\documentclass[structabstract]{aa}
\usepackage{amsmath,amssymb}
\usepackage{natbib}
\usepackage{epsfig,graphics}
\usepackage{epstopdf} 
\usepackage{rotating}
\bibpunct{(}{)}{;}{a}{}{,}

\begin{document}
\title{Exploring organic chemistry in planet-forming zones} 
\author{Jeanette E. Bast\inst{1,4}
       \and Fred Lahuis\inst{1,2}
       \and Ewine F.\ van Dishoeck\inst{1,3}
       \and Alexander G.G.M.\ Tielens\inst{1}} 

\institute{$^{1}$Leiden Observatory, Leiden University, P.O. Box 9513, 2300 RA Leiden, The Netherlands \\
$^{2}$SRON Netherlands Institute for Space Research, P.O. Box 800, 9700 AV Groningen, The Netherlands \\
$^{3}$Max Planck Institute for Extraterrestrial Physics, Giessenbachstrasse 1, 85748 Garching, Germany \\
\email{bast@strw.leidenuniv.nl}
}

\date{February 21, 2012}


\begin{abstract}
{ Over the last few years, the chemistry of molecules other than CO in
  the planet-forming zones of disks is starting to be explored with
  \textit{Spitzer} and high-resolution ground-based data. However, these
  studies have focused only on a few simple molecules.
}
{ The aim of this study is to put observational constraints on the
  presence of more complex organic and sulfur-bearing molecules
  predicted to be abundant in chemical models of disks
  and to simulate high resolution spectra in view of future missions.
}
{ High $S/N$ \textit{Spitzer} spectra at 10--30 $\mu$m of the near edge-on disks
  IRS 46 and GV Tau are used to search for mid-infrared absorption
  bands of various molecules. These disks are good laboratories
  because absorption studies do not suffer from low line/continuum
  ratios that plague emission data. Simple LTE slab models are used to
  infer column densities and excitation temperatures for detected
  lines and upper limits for non-detections.
}
{ Mid-infrared bands of HCN, C$_2$H$_2$ and CO$_2$ are clearly detected
  toward both sources. As found previously for IRS 46 by Lahuis et
  al.\ (2006), the HCN and C$_2$H$_2$ absorption arises in warm gas
  with excitation temperatures of 400--700 K, whereas the CO$_2$
  absorption originates in cooler gas of $\sim$250 K. Absolute column
  densities and their ratios are comparable for the two sources. No
  other absorption features are detected at the 3$\sigma$
  level. Column density limits of the majority of molecules predicted
  to be abundant in the inner disk --- C$_2$H$_4$, C$_2$H$_6$, C$_6$H$_6$, 
  C$_3$H$_4$, C$_4$H$_2$, CH$_3$, HNC, HC$_3$N, CH$_3$CN, NH$_3$ and SO$_2$---
  are determined and compared with disk models. 
  Simulations are also performed for future higher
  spectral resolution instruments.
}
{ The inferred abundance ratios and limits with respect to C$_2$H$_2$
  and HCN are roughly consistent with models of the chemistry in high
  temperature gas. Models of UV irradiated disk surfaces generally
  agree better with the data than pure X-ray models. The limit on
  NH$_3$/HCN implies that evaporation of NH$_3$-containing ices is
  only a minor contributor. The inferred abundances and their limits
  also compare well with those found in comets, suggesting that part
  of the cometary material may derive from warm inner disk gas.
  The high resolution simulations show that future instruments
    on JWST, ELTs, SOFIA and SPICA can probe up to an order of
    magnitude lower abundance ratios and put important new constraints
    on the models, especially if pushed to high $S/N$ ratios.
} 
\end{abstract}


\maketitle

\section{Introduction}

The chemical composition of the gas in the inner regions of
circumstellar disks plays an important role in determining the
eventual composition of the comets and atmospheres of any planets that
may form from that gas \citep[see reviews by][]{Prinn1993,
  Ehrenfreund2000,Markwick2004,Bergin2009}. In the last few years,
observations with the \textit{Spitzer} Space Telescope have revealed a
rich chemistry in the inner few AU of disks around low-mass stars,
containing high abundances of HCN, C$_2$H$_2$, CO$_2$, H$_2$O and OH \citep[][]{Lahuis2006,Carr2008,Salyk2008, Pascucci2009, Carr2011, Kruger2011, 
  Najita2010, Pontoppidan2010, Salyk2011a}. Spectrally and spatially
resolved data of CO using ground-based infrared telescopes at 4.7
$\mu$m show that the warm molecular gas is indeed associated with the
disk \citep[e.g.,][]{Najita2003, Brittain2003, Brittain2007,
  Brittain2009, Blake2004, Pontoppidan2008, Salyk2011b, Brown2012},
with in some cases an additional contribution from a disk wind
\citep{Bast2011,Pontoppidan2011}.  Spectrally resolved ground-based
observations have also been obtained of OH and H$_2$O at 3 $\mu$m
\citep{Carr2004, Mandell2008, Salyk2008, Fedele2011}, and most
recently of HCN and C$_2$H$_2$ \citep{Gibb2007, Doppmann2008, Mandell2012}.  All of these data
testify to the presence of an active high-temperature chemistry in the
upper layers of disks that drives the formation of OH, H$_2$O and
small organic molecules. However, it is currently not known whether
this gas contains more complex organic molecules which may eventually
become part of exoplanetary atmospheres.

Observations of large interstellar molecules are usually carried out using
(sub-)millimeter telescopes. A wide variety of complex organic species
have been found in low- and high-mass protostars at the stage when the
source is still embedded in a dense envelope \citep[see][for
review]{Herbst2009}. For disks, the pure rotational lines of CO, H$_2$O,
HCO$^+$, H$_2$CO, HCN, N$_2$H$^+$, CN, C$_2$H, SO, DCO$^+$ and DCN
have been reported but more complex molecules have not yet been
detected \citep[e.g.,][]{Dutrey1997,Kastner1997, Thi2004,
  Fuente2010, Henning2010, Oberg2011,Hogerheijde2011}. Although these
millimeter data have the advantage that they do not suffer from dust
extinction and can thus probe down to the midplane, current facilities
are only sensitive to the cooler gas in the outer disk ($>50$AU). Even the
Atacama Large Millimeter/submillimeter Array (ALMA) with its much
improved spatial resolution and sensitivity can only readily image
molecules at $\sim$5 AU or larger in the nearest disks.  Moreover,
ALMA cannot detect molecules without a permanent dipole moment such as
C$_2$H$_2$ and CH$_4$, which are among the most abundant species in
the inner disk. Results so far show that there is no clear correlation
between the chemistry in the inner and outer parts of
the disk \citep{Oberg2011}. The chemistry in the inner regions seems
to be sensitive to different shapes of radiation fields and the
accretion luminosities \citep{Pascucci2009, Pontoppidan2010}, but
these quantities do not seem to have an impact on the chemical
composition of the colder gas further out in the disk.

Searches for more complex molecules in the inner few AU must therefore
rely on infrared techniques. However, the strong mid-infrared
continuum implies very low line/continuum ratios for emission lines,
even at high spectral resolution. Indeed, the recent VLT-CRIRES
($R=\lambda/\Delta \lambda = 10^5$) searches in the 3 $\mu$m
atmospheric window show that lines of molecules other than CO have
line/continuum ratios of typically only a few \%, and that even
relatively simple species like CH$_4$ are not detected at the
$\sim$1\% level \citep{Mandell2012}. On the other hand, absorption
lines offer a much better chance of detecting minor species for a
variety of reasons. First, absorption occurs from the ground
vibrational level where the bulk of the population resides, so that
the signal is much less sensitive to temperature.  Another advantage
is that absorption lines are relative in strength to the continuum
whereas emission lines are absolute. So the strength of the absorption
lines relative to the continuum will stay the same in sources which
have a stronger continuum whereas the emission lines will be dominated
by the continuum. Both these advantages imply that absorption lines
are easier to detect for less abundant molecular species than emission
lines.

Detection of absorption lines requires, however, a special orientation
of the disk close to edge-on, so that the line of sight to the
continuum passes through the inner disk. Only a few disks have so far
been found with such a favorable geometry: that around Oph-IRS46
\citep{Lahuis2006}, GV Tau N \citep{Gibb2007, Doppmann2008} and DG Tau B \citep{Kruger2011}. In all cases, the mid-infrared absorption bands of HCN, C$_2$H$_2$ and CO$_2$ have
depths of 5--15\%, even at the low spectral resolution $R\approx 600$
of \textit{Spitzer}. For high $S/N>100$ spectra, detection of absorption
features of order 1\% should be feasible, providing a dynamic range of
up to an order of magnitude in abundances to search for other
molecules. With increased spectral resolution and sensitivity offered
by future mid-infrared instruments such JWST-MIRI ($R\approx$3000),
and SOFIA, SPICA and ELTs ($R\geq 50,000$), another order of dynamic
range will be opened up.

A large variety of increasingly sophisticated physico-chemical models
of the inner regions of disks exist
\citep[e.g.,][]{Willacy1998,Aikawa1999,Markwick2002,Nomura2007,
  Agundez2008,Gorti2008,Glassgold2009,Nomura2009,Willacy2009,
  Woitke2009,Kamp2010,Walsh2010,Aresu2011,Gorti2011,Heinzeller2011,
  Najita2011,Vasyunin2011,Walsh2012}.  The models differ in their
treatments of radiation fields (UV and/or X-rays), the gas heating and
resulting disk structure, dynamical processes such as accretion flows
and disk winds, grain properties and chemical networks (e.g., grain
opacities, treatment of gas-grain chemistry including H$_2$ formation
at high temperature). Some models consider only the simplest molecules
in the chemistry, others have a large chemical network but publish
primarily results for species that can be observed at millimeter
wavelengths. Only \citet{Markwick2002} list the most abundant species,
including complex molecules that do not have a dipole moment, at 1, 5
and 10 AU as obtained from vertically integrated column
densities. Since infrared observations probe only part of the disk
down to where the continuum becomes optically thick, these models may
not be representative of the surface layers. Column densities
appropriate for comparison with infrared data have been presented by
\citet{Agundez2008} and \citet{Najita2011} but do not provide data for more
complex molecules.  \citet{Woods2007} and \citet{Kress2010} consider
PAH processing in the inner disk and study its impact on the
abundances of related species like benzene and C$_2$H$_2$.  Note that
PAHs are generally not detected in disks around T Tauri stars,
including the two disks studied here, at levels a factor of 10--100
lower than found in the interstellar medium
\citep{Geers2006,Oliveira2010}.

Observations of molecules in comets provide another interesting data
set for comparison with protoplanetary disks.  Solar system comets
were likely formed at distances of about 5--30 AU in the protosolar
nebula. Many volatile molecules are now routinely observed in cometary
atmospheres at infrared and millimeter wavelengths, including species
as complex as C$_2$H$_6$, CH$_3$OH, and even (CH$_2$OH)$_2$
\citep[see][for reviews]{Mumma2011,Bockelee2011}. It is still debated
whether the abundances measured in comets directly reflect those found
in the dense envelopes around protostars or whether they result from
processing and mixing material from the inner and outer disk into the
comet-forming zone. Putting constraints on the inner disk abundances
of these molecules will be important to probe the evolution of
material from the natal protosolar nebula to the formation of icy
bodies. 

  In this study we use the existing high $S/N$ \textit{Spitzer} spectra
  of IRS 46 and GV Tau to put, for the first time, upper limits on
  various molecules in the inner disk: HNC, CH$_3$, C$_2$H$_4$,
  C$_2$H$_6$, C$_3$H$_4$, HC$_3$N, C$_6$H$_6$, NH$_3$, C$_4$H$_2$,
  CH$_3$CN, H$_2$S and SO$_2$. These molecules were selected to
  include most of the top 15 highest vertical column density molecules
  at 1 and 5 AU by \citet{Markwick2002}. The selected species can in principle 
  directly test the predictions
  of models of inner disk chemistry. The list also
  contains several molecules observed in cometary atmospheres and two more 
  molecules with a permanent dipole moment, HNC and HC$_3$N, which
  together with HCN can be observed at both infrared and millimeter
  wavelengths and can thus be used to connect the inner and outer disk
  chemistries through ALMA imaging.

  The mid-infrared spectra of IRS\,46 and GV\,Tau contain detections
  of C$_2$H$_2$, CO$_2$ and HCN which are analyzed here in terms of
  column densities and abundances, following the same strategy as for
  IRS\,46 in \citet{Lahuis2007}. The earlier detections of HCN and
  C$_2$H$_2$ toward GV\,Tau were performed in the 3 $\mu$m window with
  higher spectral resolution \citep{Gibb2007} 
but no \textit{Spitzer} results have yet been presented on
this source. We have reduced the GV\,Tau spectrum and rereduced the IRS\,46
spectrum with the latest IRS pipeline in order to allow for a consistent
comparison between the two sources. For the same reason, we have also
reanalyzed the IRS\,46 spectrum using the same analysis software.




A description of the observations and the reduction of the data for
IRS 46 and GV Tau is presented in Section \ref{observations} together
with some information about these two protoplanetary disks.  Section
\ref{spectra} presents an overview of the observed and modeled
spectra, whereas \ref{model} uses a synthetic local thermodynamic
equilibrium (LTE) model to estimate column densities and excitation
temperatures for HCN, C$_2$H$_2$ and CO$_2$ toward GV Tau and compares
the results with those for IRS 46.  Section \ref{results} presents the
upper limits for the various molecules toward the two sources and
Section \ref{future} shows how future instruments can provide more
stringent limits.  Discussion and comparison to chemical models is
performed in Section \ref{discussion}. A summary of the main
conclusions is found in Section \ref{conclusions}.

\section{Observations} \label{observations}

\subsection{IRS 46 and GV Tau} \label{sources}

The observations of IRS 46 and GV Tau were made using
\textit{Spitzer}-IRS in both the Short-High (SH; $9.9-19.6\,\mu$m) and
Long-High (LH; $18.7-37.2\,\mu$m) modes with a spectral resolving power of
$R={\lambda}/{\delta}{\lambda}=600$.

Oph-IRS\,46 was observed at $\alpha=16^h27^m29^s.4$ and
$\delta=-24^o39\arcmin16\arcsec.3$ (J2000), located in the Ophiuchus
molecular cloud at a distance of around 120 pc \citep{Loinard2008}.
IRS\,46 was initially observed in 2004 as part of the Cores to Disks
\textit{Spitzer} legacy program \citep{Evans2003} and in 2008 and 2009
at multiple epochs to search for variability.  Its mid-infrared
spectral energy distribution rises strongly with wavelength, as
expected for a near edge-on disk \citep{Crapsi2008}.  Strong HCN,
C$_2$H$_2$ and CO$_2$ absorption has been detected with \textit{Spitzer}
and attributed to arise from warm gas in the surface layers of the
inner few AU of the disk, seen in absorption against the continuum
produced by the hot inner rim on the near and far side of the star
\citep{Lahuis2006}.  \citet[][and in prep.]{Lahuis2011} show that hot
water emission lines are also detected. More interestingly, strong
variation in the depth of the molecular absorption bands as well as in
the strength of the water emission lines and the mid-IR contiuum is
observed on timescales of a few years. The data used here are the
original observations obtained on August 29, 2004 as part of AOR\#
0009829888 and published by \citet{Lahuis2006}. These show the deepest
molecular absorptions, thus providing the best upper limits of column
densities of other species relative to the observed C$_2$H$_2$, HCN
and CO$_2$ column densities.

GV\,Tau is a T\,Tauri star that is partly embedded in the L1524
molecular cloud. Its observations were positioned at
$\alpha=4^h29^m25^s.8$ and $\delta=+24^o33\arcmin00\arcsec5$ (J2000).
It has an infrared companion about 1.2$\arcsec$ to the north. The
companion is named GV Tau N and the primary optical source is called GV Tau
S. \citet{Gibb2007} detected HCN and C$_2$H$_2$ toward GV Tau N using
Keck-NIRSPEC at L-band, however, no such detections were made toward
GV Tau S.  GV Tau has subsequently been observed using the IRS SH mode
at multiple epochs with \textit{Spitzer} in a GO4 program (PI, F.Lahuis;
program ID 50532).  For the SH part of the spectrum the GO4 data (AOR
\# 0022351616, 0028247808, 0028247552 and 0031618304) were used.  For
the LH part, data from the \textit{Spitzer} GTO program observed on 
2 March 2004 as part of AOR \# 0003531008 were adopted. Note that the
\textit{Spitzer}-IRS aperture does not resolve the GV Tau binary, in
contrast to NIRSPEC. The \textit{Spitzer} spectra therefore combine
emission and absorption of GV Tau N and GV Tau S. Both GV Tau N and GV
Tau S are variable at 2\,$\mu$m \citep{Leinert2001, Koresko1999}.
This variability has been attributed to variable accretion mechanisms for
GV Tau N and variation in the extinction due to inhomogeneities in the
circumstellar material for GV Tau S \citep{Leinert2001}. However, the
multi-epoch \textit{Spitzer} data do not show significant mid-infrared
variation on timescales of a few months up to a few years, in contrast
with the near-IR ground-based results and with IRS 46. 
The mid-IR continuum of GV Tau N is about an order of magnitude
brighter than the continuum of GV Tau S between $8-13$\,$\mu$m
\citep{Przygodda2004, Roccatagliata2011}. 
Since no absorption was seen in GV Tau S in \citet{Gibb2007} and
\citet{Doppmann2008} it is assumed that the majority of the absorption
arises toward GV Tau N. However, the continuum emission from the
southern source captured in the \textit{Spitzer}-IRS aperture slightly
reduces the total optical depth of the absorption lines in the
spectrum. To put an upper limit on the added uncertainty caused by the
additional continuum emission from GV Tau S, it is assumed that the
mid-IR continuum of GV Tau N is $\sim10$
times stronger than
that of GV Tau S, resulting in an additional uncertainty of 
$\sim 1-2$\,\% 
for features that are $\sim5-20$\,\% deep. Since the effect
is minor, no correction is made for the column densities derived here.



\subsection{Data reduction}

The data reduction started with the BCD (Basic Calibrated Data) images
from the \textit{Spitzer} archive processed through S18 pipeline. The BCD
images were then processed using the Cores to Disks (c2d) analysis
pipeline \citep{Lahuis2006b, Kessler2006}. The main processing steps
are background correction, bad-pixel removal, spectral extraction,
defringing, order matching and spectral averaging. Two extraction
methods were used; 1) a full aperture extraction from the BCD images
and 2) an optimal extraction using an analytical psf
\citep{Lahuis2007} defined using a set of high $S/N$ calibration
stars. For both extractions a relative spectral response function
(RSRF) calibration is applied with $\xi$ Dra as the main reference star
using MARCS models taken from the \textit{Spitzer} science center
\citep{Decin2004}.

For all observations the extraction method giving the best $S/N$ was
used to produce the final spectra. The two (partly) independent
extraction methods further allow to better discriminate between
artifacts and true science features.

\section{Results}

\subsection{Spectra}
\label{spectra}

 \begin{figure*}
\centering
{
 \includegraphics[width=160mm, angle=0.0]{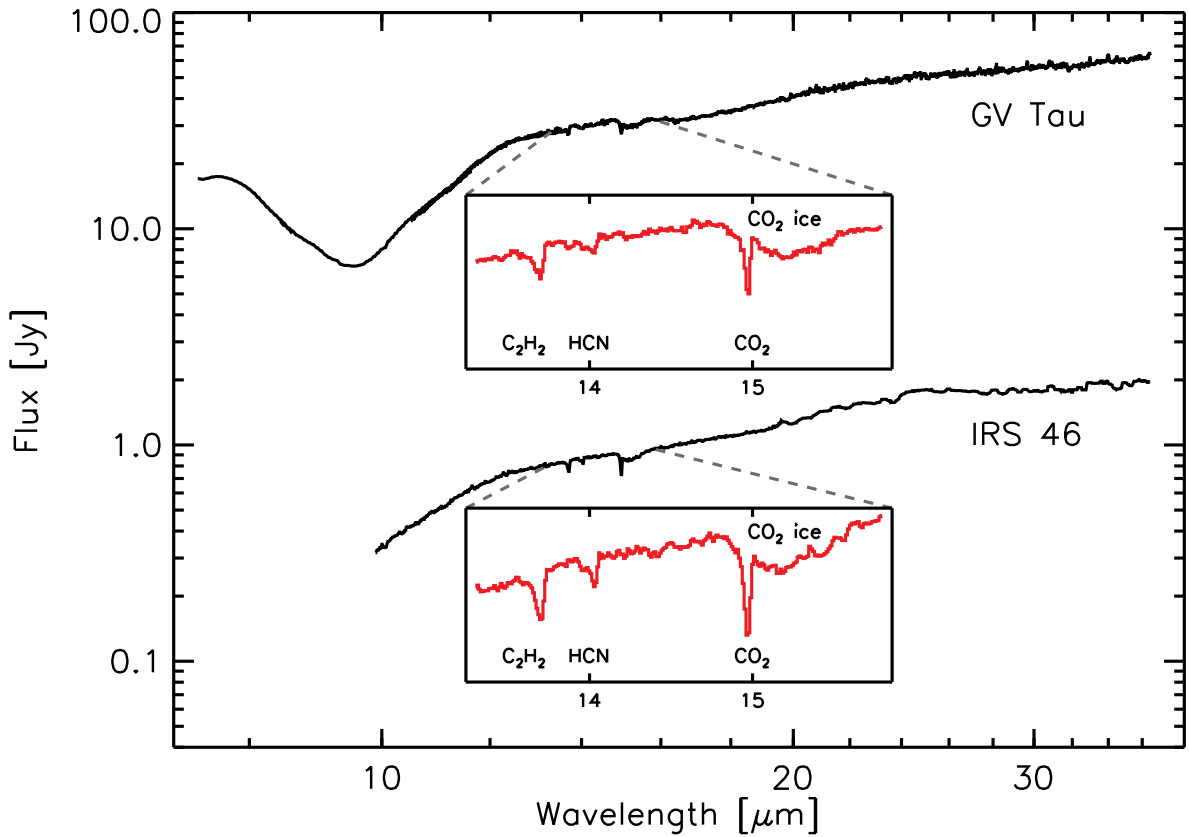}
 \caption{Spectra of the protoplanetary disks around IRS 46 and GV Tau 
taken with the \textit{Spitzer}-IRS.}
 \label{fig:full_spectra}
}
\end{figure*} 

Figure \ref{fig:full_spectra} shows the spectra of IRS\,46 and GV\,Tau
over the $10-37\,\mu$m region.
Both spectra differ by more than an order of magnitude in
the mid-IR continuum level, but they show a striking resemblance
in spectral shape and spectral features.  
The absorption bands of gaseous
C$_2$H$_2$ $\nu_5$, HCN $\nu_2$ and CO$_2$ $\nu_2$ can be clearly seen
at 13.7, 14.0 and 15.0 $\mu$m, together with the solid CO$_2$ feature
at $15-16\,\mu$m. To search for other molecules, a local continuum has
been fitted to the broad spectral features and divided out. The $S/N$
on the continuum is typically 100 or better. No other obvious
absorption features are detected at the few\,\% absorption level.
The model spectra with derived column densities and upper limits are
described below.

\subsection{C$_2$H$_2$, HCN and CO$_2$}
\label{model}

To extract quantitative information from the spectra, a simple local
thermodynamical equilibrium (LTE) absorption slab model has been used
to fit the data. The free parameters in the model are the excitation
temperature, the integrated column density along the line of sight and
the intrinsic line width, characterized by the Doppler $b$-value. The
excitation temperature sets the level populations of the molecule
using the Boltzmann distribution. The lack of collisional rate
coefficients for many of the species considered here prevents non-LTE
analyses. The model spectrum is convolved with the spectral resolution
of the instrument and resampled to the observed spectra.
More details about the model and the molecular parameters and data
that are used for the three detected molecules 
can be found in \citet{Lahuis2000,Lahuis2007,Boonman2003} and
in Table~1.

\begin{table*}
\caption{Basic molecular data}
\label{tab:moldata}
\begin{minipage}[t]{\textwidth}
\renewcommand{\footnoterule}{}  
\thispagestyle{empty}
\begin{tabular}{l l l c c c c}
\hline
\hline
{\Large\strut} Molecule & Formula & Band 
& $\tilde \nu$\footnote{Central wavenumber of band and band strengths from \textit{Constants for molecules of 
  astropysical interest in the gas phase} by J. Crovisier \newline
  http://wwwusr.obspm.fr/$\sim$crovisie/basemole/}
& $S_{lit}$$^a$
& $S_{int}$\footnote{Band strength of the simulated spectra.
  $S=N_L \times \int\!\tau(\nu)\delta\nu\,/\,n$, \citep[see App. A.1][]{Helmich1996}. Calculations were performed for $T=298$\,K, $b=20$\,km\,s$^{-1}$ and 
  $n=1\cdot10^{15}$\,cm$^{-2}$ at a resolution of 3000 to keep all bands far from saturation}
& Source\footnote{H08: HITRAN 2008 \citep{Rothman2009}, 
                  G09: GEISA 2009 \citep{Jacquinet2011} and 
                  FPH: \citet{Helmich1996}} \\ 
{\Large\strut}   &         &      & [cm$^{-1}$] & [atm$^{-1}$\,cm$^{-2}$] & [atm$^{-1}$\,cm$^{-2}$] & \\
\hline
Acetylene                       & C$_2$H$_2$  & $\nu_5$ CH bending           & 729.1 &  630 &  816 & H08 \\
Carbon Dioxide                  & CO$_2$      & $\nu_2$ bending              & 667.4 &  200 &  249 & H08 \\
Hydrogen Cyanide                & HCN         & $\nu_2$ bending              & 713.5 &  257 &  286 & H08 \\
Hydrogen Isocyanide             & HNC         & $\nu_2$ bending              & 464.2 & 1570 &  798 & G09 \\
Methyl Radical                  & CH$_3$      & $\nu_2$ out-of-plane bending & 606.5 &  611 &  616 & FPH \\
Ethylene                        & C$_2$H$_4$  & $\nu_7$ CH2 waggling         & 949.2 &  324 &  320 & G09 \\
Ammonia                         & NH$_3$      & $\nu_2$ symmetric bendng     & 950.0 &  568 &  614 & H08 \\
Sulphur Dioxide                 & SO$_2$      & $\nu_2$ bending              & 517.6 &  113 &   97 & G09 \\
Ethane                          & C$_2$H$_6$  & $\nu_9$ CH3 rocking          & 822.0 &   36 &   29 & H08 \\
Diacetylene / Butadiyne         & C$_4$H$_2$  & $\nu_8$ CH bending           & 627.9 &  437 &  229 & G09 \\
Benzene                         & C$_6$H$_6$  & $\nu_4$ CH bending           & 673.5 &  250 &  212 & G09 \\
Propyne / Methyl Acetylene      & C$_3$H$_4$  & $\nu_9$ CH bending           & 638.6 &  360 &  201 & G09 \\
Cyanoacetylene / Propynenitrile & HC$_3$N     & $\nu_5$                      & 663.4 &  278 &   94 & G09 \\
Methyl Cyanide / Acetonitrile   & CH$_3$CN    & $\nu_4$ CC stretch           & 920.3 &    6 &    3 & G09 \\
\hline
\end{tabular} 
\label{tab:up_limits}
\end{minipage}
\end{table*}


 \begin{figure*}
\centering
{ 
 \includegraphics[width=170mm, angle=0.0]{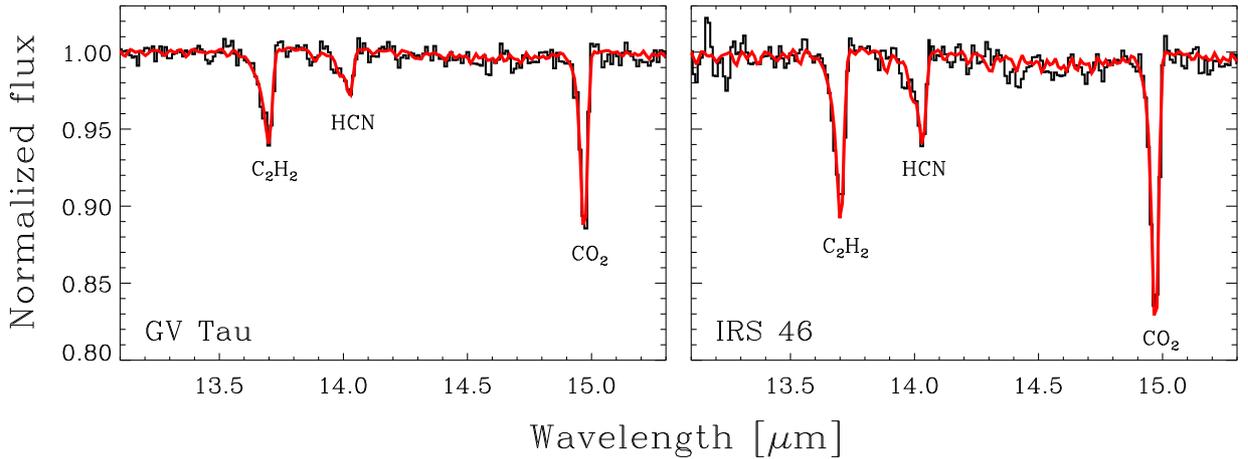}
 \caption{Continuum normalized spectra of GV\,Tau and IRS\,46. 
 Plotted in black are the observed spectra and overplotted
   in red the best-fit synthetic spectra to the absorption bands of
   C$_2$H$_2$ $\nu_5$=1--0, HCN $\nu_2$=1--0 and CO$_2$
   $\nu_2$=1--0. See Table~\ref{tab:gvtau_irs46} for best fit model
   parameters.}
 \label{fig:spectral_fits}
}
\end{figure*}

Figure \ref{fig:spectral_fits} presents a blow-up of the $13-15\,\mu$m
range of the GV\,Tau and IRS\,46 spectra with the continuum divided
out. Included are the best-fitting model spectra. The figure clearly
shows that the $P$- and $R$-branch lines are difficult to detect at
the \textit{Spitzer}-IRS spectral resolution of $R=600$, however the
$Q$-branches of C$_2$H$_2$, HCN and CO$_2$ are easily seen.  In
addition the $Q$-branch changes its form and depth with excitation
temperature and column density. In particular, the depth of the
$Q$-branch decreases with increased excitation temperature for the
same total column density and broadens to the blue side due to an
increase in the population of the higher rotational levels. A higher
column density on the other hand increases the central depth of the
$Q$-branch since more molecules absorb photons. As can be seen in
\citet{Lahuis2000} the $Q$-branch is sensitive to the adopted Doppler
$b$-value with the magnitude of the effect depending on the
temperature and column density.
%
%
Spectrally resolved data obtained with Keck-NIRSPEC \citep{Salyk2011b,Lahuis2006}
and within our VLT-CRIRES survey \citep{Pontoppidan2011,Brown2012}
show that the HCN and CO lines have $b\approx 12$ km s$^{-1}$.  In our
analysis we therefore adopt a Doppler $b$-value of 10 km s$^{-1}$.
\citet{Lahuis2006} used $b$=5 km s$^{-1}$ 
for IRS\,46 which gives slightly increased temperatures and reduced
column densities by $\sim 5-10$\,\%.

  \begin{table}
\caption{Results from molecular fits to GV\,Tau and IRS\,46 absorption features}
\medskip
 \begin{minipage}[t]{\columnwidth}
 \renewcommand{\footnoterule}{}  
 \centering
  \thispagestyle{empty}
  \begin{tabular}{l c c c c}
  \hline
   \hline
{\large\strut}& \multicolumn{2}{c}{Temperature [K]} & \multicolumn{2}{c}{Column density [10$^{16}$ cm$^{-2}$]} \\
{\large\strut}Source & IRS 46 & GV Tau & IRS 46 & GV Tau \\
\hline
{\large\strut}C$_2$H$_2$ & $490\,\pm\,^{50}_{30}$ & $720\,\pm\,^{60}_{40}$ & 2.1 $\pm$ 0.4  & 1.4 $\pm$ 0.3 \\
{\large\strut}HCN        & $420\,\pm\,^{40}_{25}$ & $440\,\pm\,^{40}_{30}$ & 3.7 $\pm$ 0.8  & 1.8 $\pm$ 0.4 \\
{\large\strut}CO$_2$     & $250\,\pm\,^{25}_{15}$ & $250\,\pm\,^{25}_{15}$ & 8.4 $\pm$ 1.1  & 5.1 $\pm$ 0.7 \\
\hline  
\end{tabular} 
 \end{minipage}
\label{tab:gvtau_irs46}
\end{table}

A grid of synthetic spectra of C$_2$H$_2$, HCN and CO$_2$ was made for
a range of column densities and temperatures and fitted to the data
obtained for IRS\,46 and GV Tau. The best fit as presented in
Fig.~\ref{fig:spectral_fits} was determined by finding the minimum
difference between data and model as measured by the $\chi^2$
values. The derived column densities and excitation temperatures for
the different molecules are summarized in Table
\ref{tab:gvtau_irs46}. For IRS 46, the values are consistent with
those of \citet{Lahuis2006} taking into account that a higher
$b-$value was adopted and that the final spectrum presented in this
paper has a slightly higher $S/N$ ratio compared to \citet{Lahuis2006}.  
It is seen that the temperatures of the
different molecules and their column density ratios are comparable
between GV Tau and IRS 46. This supports the hypotheses that both
sources are inclined disks with similar characteristics. In both
sources CO$_2$ has the highest column density but the lowest
temperature, whereas the HCN/C$_2$H$_2$ ratio is slightly above
unity. In their 2 -- 5 $\mu$m study, \citet{Gibb2007} however find
significantly lower temperatures for C$_2$H$_2$ (170 $\pm$ 20 K) and
HCN (115 $\pm$ 20 K) compared to our estimated temperatures of about
400 to 700\,K. However in later Keck-NIRSPEC L-band observations,
\citet{Doppmann2008} and \citet{Gibb2011} detect lines out to much higher $J$
values, indicating warmer gas around 500 K. Our mid-infrared results
are therefore not inconsistent with the near-IR data.

The high resolution near-IR data for IRS 46 and GV Tau N show that the
spectral lines are shifted in velocity. For IRS46, CO and HCN are
blueshifted by about 24 km s$^{?1}$ with respect to the cloud
\citep{Lahuis2006} whereas for GV Tau the HCN lines are redshifted by
about 13 km s$^{-1}$ compared with the star \citep{Doppmann2008}. This
could indicate that the observed HCN, C$_2$H$_2$ and CO$_2$ absorption
originates in a disk wind or infalling envelope rather than the disk
itself.  However, the high densities needed to excite these higher
$J$-transitions as well as the constraints on the size of the high
abundance region ($<$11 AU) imply that the absorption lines have an
origin in outflowing or infalling gas that must be very closely
related to the disk itself with a chemistry similar to that of the
disk. This is further discussed in \citet{Lahuis2006, Gibb2007,
  Doppmann2008, Kruger2011} and
\citet{Mandell2012}. \citet{Fuente2012} recently imaged the warm HCN
associated with the disk of GV Tau N at millimeter wavelengths and
found an emitting radius of less than 12 AU.

\subsection{Other molecules}
\label{results}

\subsubsection{Overview and molecular data}

Table \ref{tab:moldata} summarizes the molecular data
(vibrational mode, line positions, 
band strengths) used for all molecules for which searches have been
made toward IRS 46 and GV Tau in the 10--30 $\mu$m wavelength range,
together with the main references from which they have been
extracted. Only the intrinsically strongest bands of each molecule
have been chosen; weaker bands are ignored. Note that isotopologues
and vibrationally excited states or `hot bands' are not included in
the data sets, except in the fitting of the observed spectra of HCN,
CO$_2$ and C$_2$H$_2$.  Hot bands are expected to be suppressed in
full non-LTE calculations, where the excitation of the higher
vibrational levels is subthermal at densities below $\approx 10^{10}$
cm$^{-3}$.  Synthetic spectra are generated following the procedures
as described in \citet{Helmich1996} and \citet{Lahuis2000}.

Our spectra use molecular data from various databases listed in Table
\ref{tab:moldata}. To assess their reliability, we have computed the
integrated absorption band strengths of the simulated spectra and
compared them with independent band strengths tabulated in the
literature (columns 5 and 6 of Table \ref{tab:moldata}). This table
shows that the band strengths of all species agree to within factor of
$\sim3$ and in most cases much better.  For the purpose of this paper
this accuracy is sufficient and differences between databases and
other literature values are not pursued here. It should be noted,
however, that there are (sometimes significant) differences between
line lists and not all line lists are complete.

 \begin{figure*}
\centering
{
 \includegraphics[width=170mm, angle=0.0]{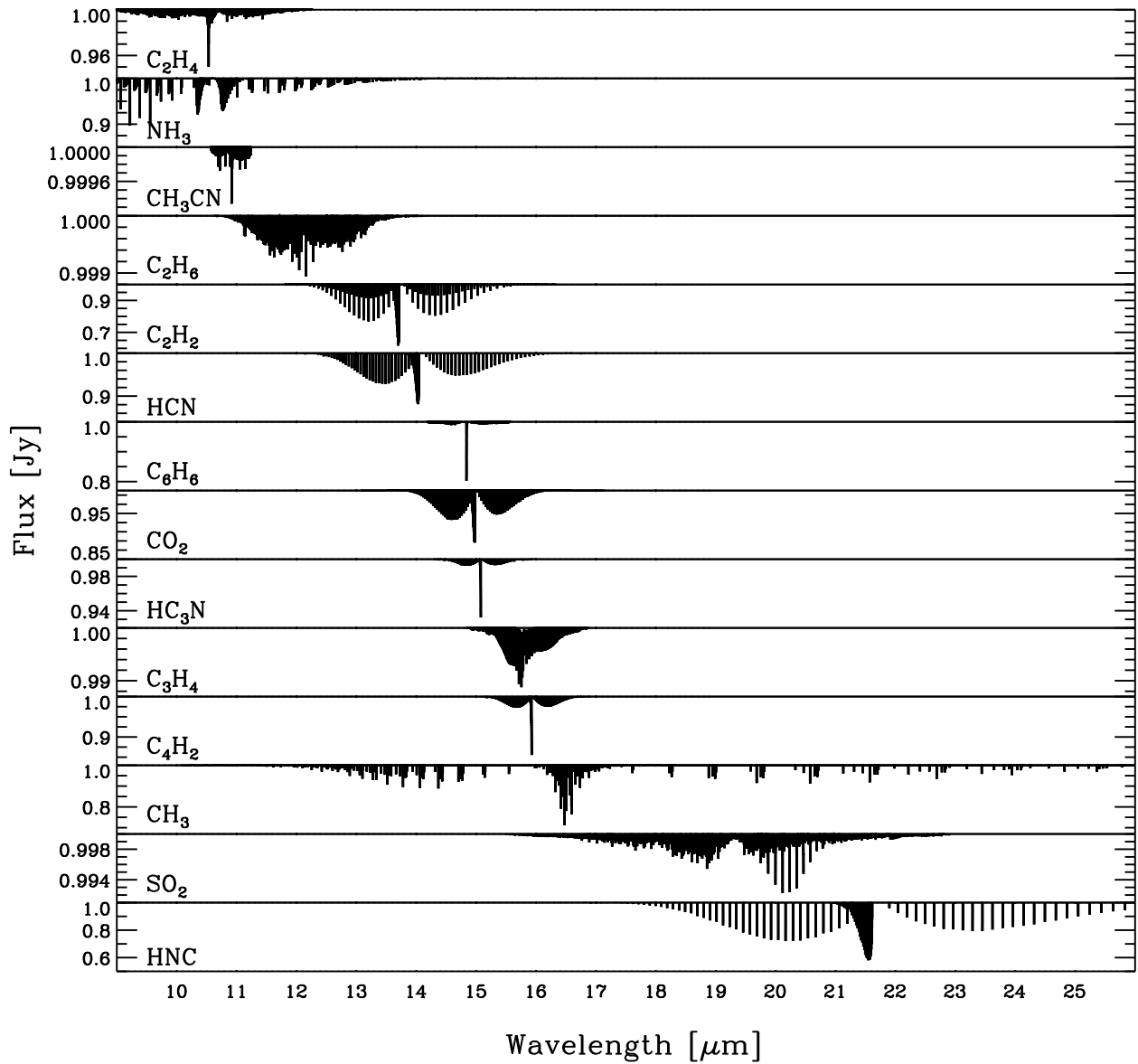}
 \caption{Synthetic spectra of the different molecules for $T_{\rm ex}$=500 K 
and for a resolving power of $R$=50,000. 
The column density of each molecule is set to be 1$^.$10$^{16}$ cm$^{-2}$ which is the same as in Fig.~\ref{fig:c2h2_plot_tot}--\ref{fig:hc3n_plot_tot}}
 \label{fig:synth_overview}
}
\end{figure*} 

Figure~\ref{fig:synth_overview} presents an overview of the simulated
LTE spectra for all molecules considered here at $T=500$~K, $R=50,000$
and a column density of 1$^.$10$^{16}$ cm$^{-2}$.  A variety of
absorption patterns is seen, depending on the characteristics and
symmetries of the individual bands involved. Together, they span most
of the wavelength range observed by \textit{Spitzer}.


\subsubsection{Upper limits from \textit{Spitzer} data}

\begin{figure*}
\centering
{
 \includegraphics[width=160mm, angle=0.0]{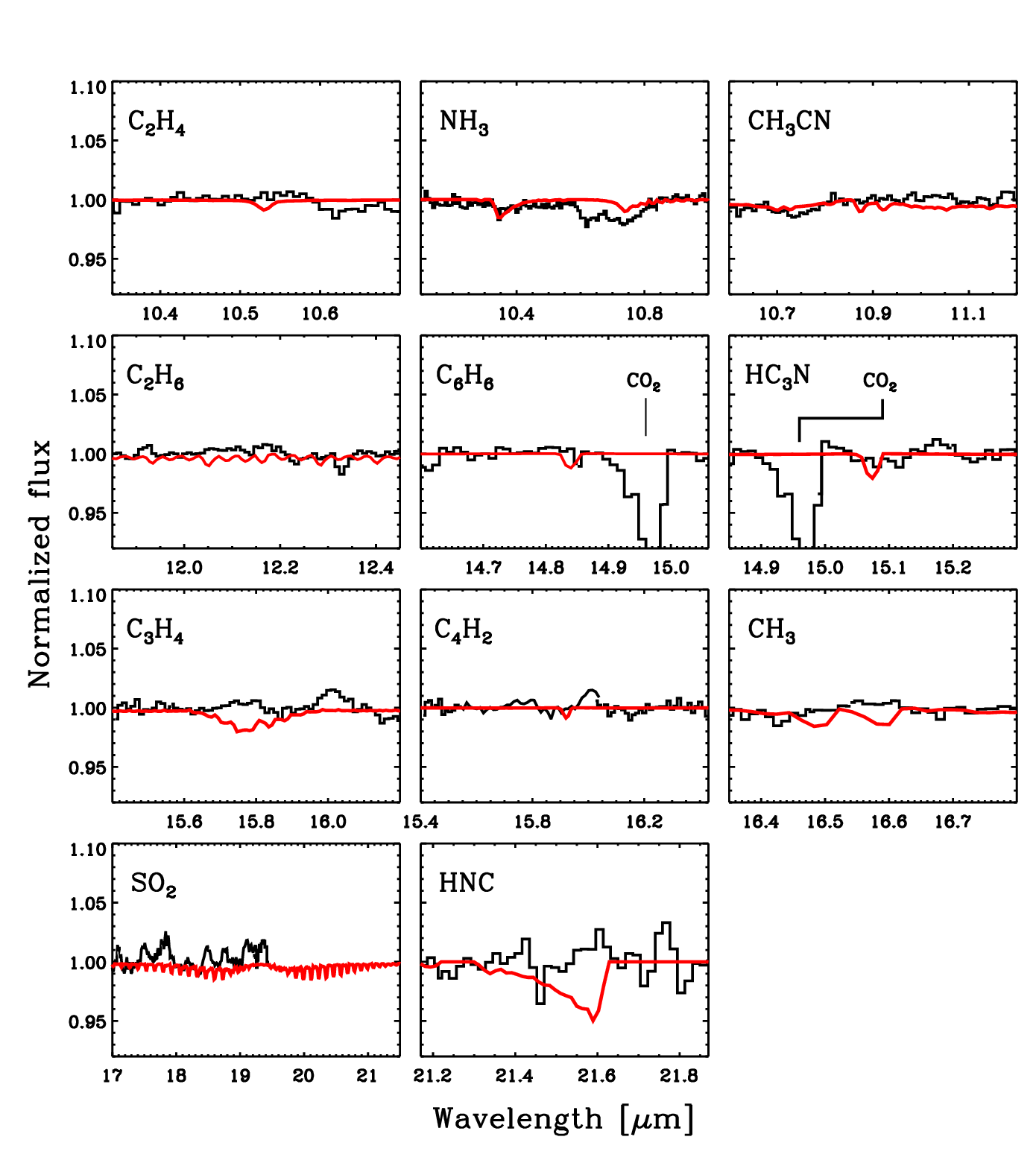}
 \caption{Blow-ups of synthetic spectra (in red) for different molecules at a
   3$\sigma$ maximum optical depth compared with the observed spectrum
   of GV Tau (in black).}
 \label{fig:gvtau_up}
}
\end{figure*}

For all species without detections upper limits were derived
  from the local normalized spectrum based on the rms noise over a
  region of 10 resolution elements. The definition of the continuum
  determines the derived rms noise and therefore the upper limits.
  To be conservative, the local continuum was defined by a straight
  line fit to clean regions either side of the spectral feature.
  The upper limit on the column density was 
  taken to be the value for which the synthetic spectrum has a feature
  depth of 3$\sigma$. Examples for the case of GV Tau are presented
  in Fig.~\ref{fig:gvtau_up}.
  For some species, e.g. CH$_3$CN, the model lies below the 
  continuum in the zoomed-in region presented in Fig.~\ref{fig:gvtau_up}
  due to blending of weak $P$- and $R$-branch lines at the 
  \textit{Spitzer} resolution. Over the wider wavelength range on
  which the continuum and model are defined both reach unity.

The depth of a molecular feature depends on the column
  density, the temperature and the adopted Doppler parameter $b$. In
  general, the features become stronger with increasing column density
  and decreasing temperature.  For high column densities, the
  transitions become optically thick and the corresponding absorption
  lines saturate.  The absorption depth then no longer increases
  linearly with column depth, with the saturation being stronger for
  smaller $b$ values.

\begin{figure*}
\centering
{
 \includegraphics[width=160mm, angle=0.0]{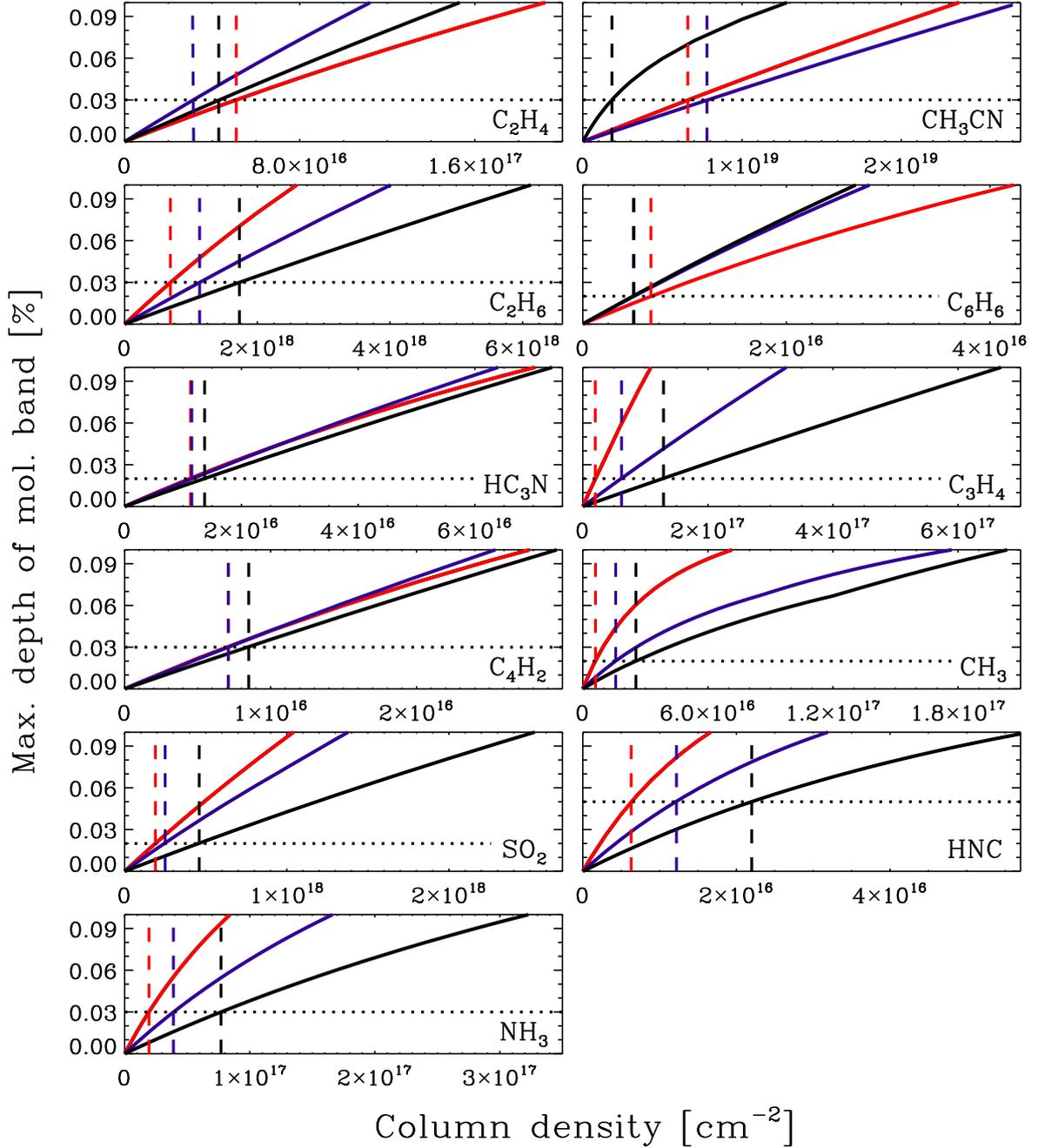}
 \caption{Variation of the maximum depth of the spectral features
as a function of column density at excitation temperatures of 200 K (red), 
500 K (blue) and 1000 K (black). 
The 3$\sigma$ observational limit for IRS 46 is marked with a black dotted line 
and the colored dashed lines show the corresponding upper limits
on the column densities for the three temperatures.}
 \label{fig:col_peak1}
}
\end{figure*}

Fig.~\ref{fig:col_peak1} shows how the relative intensity of
  the spectral features change with temperature and column density for
  each molecule at the \textit{Spitzer} resolution and
  $b=5$\,km\,s$^{-1}$ (we adopt a moderately low value for $b$ to
  illustrate the effects of saturation). The effect of saturation is
  clearly seen in some of the spectra, e.g. CH$_3$ at 200\,K, 
  where the curves deviate from linear at higher column densities.
  For reference, Figures \ref{fig:c2h2_plot_tot} --
  \ref{fig:ch3cn_plot_tot} in the Appendix present an overview of the
  simulated LTE spectra at $R=600$ for all molecules considered here
  at additional temperatures of 200 and 1000\,K.  Not all molecules
  follow the expected trend of depth versus column density and
  temperature. This can be explained when looking in detail at the
  different spectra.  For example, HNC follows the expected trend of
  decreasing depth with increasing temperature
  (Fig.~\ref{fig:hnc_plot_tot}), but not C$_6$H$_6$
  (Fig.~\ref{fig:c6h6_plot_tot}). The latter behavior is due to the
  low spectral resolving power $R=600$ which does not resolve the
  intrinsically narrow $Q-$branch of this heavy molecule.  At higher
  resolving power, however, the strength of $Q$-branch does in fact
  decrease with higher temperature as expected. 

Fig.~\ref{fig:col_peak1} includes the 3$\sigma$ absorption
  depth limits for IRS\,46 as dotted lines and the corresponding
  upper limits on the column densities for 200, 500 and 1000\,K 
  as dashed lines.  
  Saturation is not significant for any of the
  molecules so we can adopt these upper limits for IRS\,46 and GV\,Tau
  to compare with C$_2$H$_2$ and HCN derived with
  $b=10$\,km\,s$^{-1}$.  The derived 3$\sigma$ upper limits on the
  column densities are presented in Table~\ref{tab:up_limits_irs46} and
  \ref{tab:up_limits_gvtau}. 

\begin{figure*}
\centering
{
 \includegraphics[width=150mm, angle=0.0]{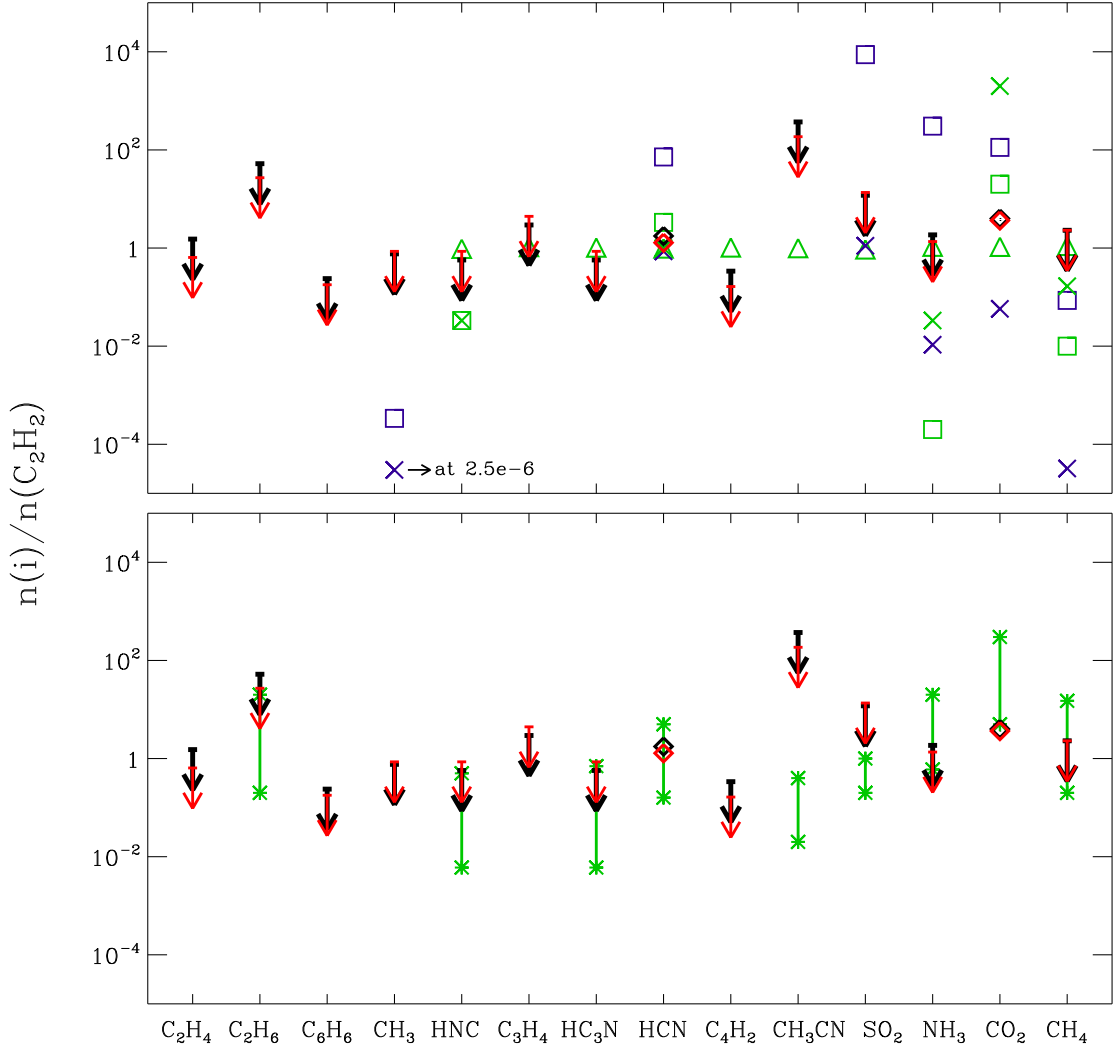}
 \caption{Comparison of observed inner disk abundance ratios
   (wrt. C$_2$H$_2$) with chemical models (upper panel) and 
   cometary observations (lower panel) for IRS\,46 and GV\,Tau. 
   In both panels detections (diamonds) and upper limits (arrows) 
   for IRS 46 (black) and GV Tau (red) are indicated.
   Upper panel: Abundance ratios in the disk model by \citet{Markwick2002}
   at 5 AU (green triangle), from the reference disk model at 1 AU by
   \citet{Najita2011} (blue square) and at a O/C = 1 (blue cross), and
   from the disk model by \citet{Agundez2008} at 1 AU (green square)
   and 3 AU (green cross).  Lower panel: Observed range of cometary
   abundance ratios from \citet{Mumma2011} (green bar and stars). 
   }
 \label{fig:comet_disk_c2h2}
}
\end{figure*}

\begin{figure*}
\centering
{
 \includegraphics[width=150mm, angle=0.0]{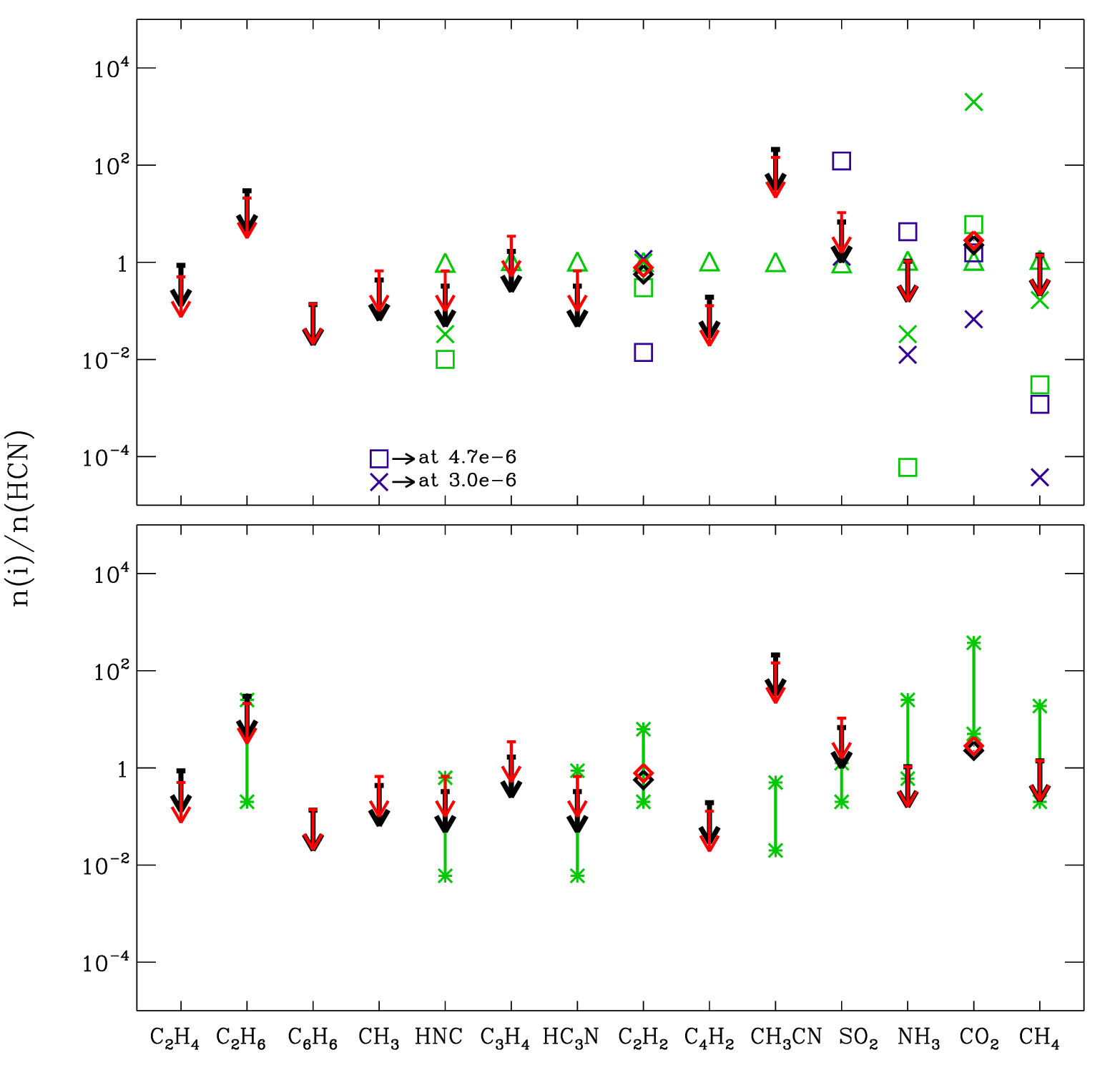}
 \caption{Comparison of observed inner disk abundance ratios
   (wrt. HCN) with chemical models (upper panel) and 
   cometary observations (lower panel) for IRS\,46 and GV\,Tau.
   Use of symbols and references identical to Figure \ref{fig:comet_disk_c2h2}.
}
 \label{fig:comet_disk_hcn}
}
\end{figure*}

Tables~\ref{tab:up_limit_c2h2} and \ref{tab:up_limit_hcn} present the
upper limits on the column densities relative to C$_2$H$_2$ and HCN,
respectively, for both IRS 46 and GV Tau. Abundance ratios relative to
C$_2$H$_2$ and HCN are typically of order unity, except for CH$_3$CN
and C$_2$H$_6$ which have particularly low band strengths (see Table
\ref{tab:moldata}). The most stringent ratios of
$<0.2-0.5$ are obtained for C$_4$H$_2$ and C$_6$H$_6$.  The ratios
are also graphically displayed in Figures \ref{fig:comet_disk_c2h2}
and \ref{fig:comet_disk_hcn} where they are compared with model
results. These results will be further discussed in \S
\ref{discussion}.



%


\subsection{High resolution spectra}
\label{future}

Figures \ref{fig:c2h2_plot_tot}, \ref{fig:co2_plot_tot} and
\ref{fig:hcn_plot_tot} in the Appendix present C$_2$H$_2$, HCN and
CO$_2$ absorption spectra at higher spectral resolution for
$R\approx 3000$, as appropriate for the JWST-MIRI instrument, and at
$R\approx 50,000$, as typical for future mid-infrared spectrometers on
an Extremely Large Telescope (ELT).
The latter spectrum is also characteristic (within a factor of 2) of
the spectral resolution of $R=100,000$ of EXES on SOFIA
\citep{Richter2006} or a potential high resolution spectrometer on the
SPICA mission \citep{Goicoechea2011}. As expected, the central $Q$
branch becomes deeper with higher spectral resolution and the $P-$ and
$R$-branches become readily detectable, allowing a more accurate model
fit to the data. The inferred column density from such data should not
change beyond the error bars derived from the low resolution data,
however.

Figures \ref{fig:hnc_plot_tot} -- \ref{fig:c3h4_plot_tot} present the
higher resolution spectra of all other molecules at $R$=3000 and
50,000 at $T_{\rm ex}=200, 500$ and 1000~K using $b=5$ km s$^{-1}$ and
a column density of 1$^.$10$^{16}$ cm$^{-2}$.  The improved
detectability of the molecules at higher spectral resolution is
obvious.  To illustrate the improvements compared with the
  current \textit{Spitzer}-IRS observations, the upper limits for the
  high resolution cases are included in Table \ref{tab:up_limits_irs46} for
IRS 46 assuming the same \textit{S/N} values.
For GV Tau, the limits at higher resolution scale similarly.
The column density limits are lower by factors of 2--10. They do not
decrease linearly with increasing resolving power, however, because
the strong $Q-$branches used to set the limits are blends of many
lines at low resolution which become separated at higher resolving
power. In some cases (C$_2$H$_6$, CH$_3$CN), the gain is very small
because the molecule absorbs less than 0.1\% of the continuum for the
adopted column density, which remains undetectable at $S/N$=100 even
at high spectral resolution.

Tables~\ref{tab:up_limit_c2h2} and \ref{tab:up_limit_hcn} 
include the column density ratios with respect to C$_2$H$_2$ and HCN
at higher resolving power. Since the C$_2$H$_2$ and HCN column
densities remain the same, the abundance ratio limits are now up to an
order of magnitude lower, thus bringing the limits in a more
interesting regime where they provide more stringent tests of chemical
models. To push the abundance ratios even lower, a higher
  $S/N$ than 100 on the continuum is needed.

Note that these limits do not take the transmission of the Earth's
atmosphere into account but assume that the strong $Q$-branches can be
observed unobscured. For ground-based instruments and to a lesser
  extent SOFIA, this is often not the case and detectability depends
both on the transmission and on the radial velocity shifts of the
sources with respect to atmospheric lines \citep{Lacy1989}.
For JWST, the better stability of a space-based platform as
well as the large mirror and the absence of telluric absorption bring
significantly higher S/N ratios than 100 within reach once the instrument
performance is well characterized in orbit.



  \begin{table*}
\caption{Inferred upper limits of column densities [$10^{16}$
    cm$^{-2}$] toward IRS 46 at different excitation temperatures and
  spectral resolving powers.
  } 
 \begin{minipage}[t]{\columnwidth}
 \renewcommand{\footnoterule}{}  
 \centering
  \thispagestyle{empty}
  \begin{tabular}{l c c c c c c}
  \hline
   \hline
{\Large\strut} & C$_2$H$_4$ & C$_2$H$_6$ & C$_6$H$_6$ & CH$_3$ & HNC & C$_3$H$_4$ \\
 \hline
{\Large\strut}$R$=600, 200 K & $<$5.1 & $<$69 & $<$0.7 & $<$0.6 & $<$0.6 & $<$2.0 \\
\ \ \ \ \ \ \ \ \ \ \ 500 K & $<$3.2 & $<$110 & $<$0.5 & $<$1.6 & $<$1.2 & $<$6.2 \\
\ \ \ \ \ \ \ \ \ \ 1000 K & $<$4.3 & $<$170 & $<$0.5 & $<$2.6 & $<$2.2 & $<$13 \\
\textit{$R$=3000, 500 K}\footnote{Upper limits at spectral resolving powers of $R = 3000$ and
50,000 at $T=500$~K adopting $S/N=100$, similar to the current
{Spitzer}-IRS data at lower spectral resolution.} & 
 \textit{$<$1.8} & \textit{$<$50} & \textit{$<$0.2} & \textit{$<$0.4} & \textit{$<$0.9} & \textit{$<$3.1} \\
 \textit{$R$=50,000, 500 K}$^a$ & \textit{$<$0.6} & \textit{$<$29} & \textit{$<$0.09} & \textit{$<$0.06} & \textit{$<$0.09} & \textit{$<$1.8} \\
\hline
 3$\sigma$ [\%]\footnote{The 3$\sigma$ limit in \% absorption at the location of the
molecular band as measured in the \textit{Spitzer}-IRS data.}  & 3 & 3 & 2 & 2 & 5 & 2 \\
 \hline
 \hline
{\Large\strut}  & HC$_3$N & C$_4$H$_2$ & CH$_3$CN & SO$_2$ & NH$_3$ \\
 \hline  
{\Large\strut}$R$=600, 200 K & $<$1.1& $<$0.7 & $<$660 & $<$19 & $<$2.0 & \\
   \ \ \ \ \ \ \ \ \ \ \  500 K & $<$1.2 & $<$0.7& $<$780 & $<$25 & $<$3.9 & \\
   \ \ \ \ \ \ \ \ \ \ 1000 K & $<$1.4 & $<$0.9 & $<$180 &  $<$46 & $<$7.7 \\
  \textit{R = 3000, 500 K} & \textit{$<$0.4} & \textit{$<$0.3} & \textit{$<$310} &  \textit{$<$8.5} & \textit{$<$1.4} \\
    \textit{R = 50,000, 500 K} & \textit{$<$0.3} & \textit{$<$0.2} & \textit{$<$46} &  \textit{$<$2.6} & \textit{$<$0.3} \\
       \hline
 3$\sigma$ [\%] & 2 & 3 &  3 &  2 &  3 & \\
 \hline
 \end{tabular} 
 \end{minipage}
\label{tab:up_limits_irs46}
\end{table*}

  \begin{table*}
\caption{Inferred upper limits of column densities [$10^{16}$ cm$^{-2}$] toward GV Tau at different excitation temperatures at a resolving power of $R$=600. 
}
 \begin{minipage}[t]{\columnwidth}
 \renewcommand{\footnoterule}{}  
 \centering
  \thispagestyle{empty}
  \begin{tabular}{l c c c c c c}
  \hline
   \hline
{\Large\strut} & C$_2$H$_4$ & C$_2$H$_6$ & C$_6$H$_6$ & CH$_3$ & HNC & C$_3$H$_4$ \\
 \hline
{\Large\strut}200 K & $<$1.2 & $<$23 & $<$0.3 & $<$0.4 & $<$0.6 & $<$2.0 \\
 500 K & $<$0.9 & $<$38 & $<$0.3 & $<$1.2 & $<$1.2 & $<$6.2 \\
1000 K & $<$0.9 & $<$57 & $<$0.3 & $<$1.8 & $<$2.2 & $<$13 \\
\hline
 3$\sigma$ [\%] \footnote{The actual 3$\sigma$ limit in \% absorption at the location of the
molecular band.}
 & 1 & 1 & 1 & 1.5 & 5 & 2 \\
 \hline
 \hline
{\Large\strut}  & HC$_3$N & C$_4$H$_2$ & CH$_3$CN & SO$_2$ & NH$_3$ \\
 \hline  
{\Large\strut}200 K & $<$1.1 & $<$0.2 & $<$230 & $<$14 & $<$0.9 & \\
   500 K & $<$1.2 & $<$0.2 & $<$260 & $<$19 & $<$1.9 & \\
 1000 K & $<$1.4 & $<$0.3 & $<$50 &  $<$34 & $<$3.7 \\
       \hline
 3$\sigma$ [\%] & 2 & 1 & 1 &  1.5 &  1.5 & \\
 \hline
\end{tabular} 
 \end{minipage}
\label{tab:up_limits_gvtau}

\end{table*}

 \begin{table*}
\caption{Observed molecular column density ratios relative to C$_2$H$_2$ in disks 
compared with different chemical models and cometary observations.} 

\begin{minipage}[t]{\columnwidth}
 \renewcommand{\footnoterule}{}  
 \centering
  \thispagestyle{empty}
  \begin{tabular}{l l l l l l l l l}
  \hline
   \hline
{\Large\strut}Ratio rel to C$_2$H$_2$ &  C$_2$H$_4$ & C$_2$H$_6$ & C$_6$H$_6$ & CH$_3$ & HNC & C$_3$H$_4$ & HC$_3$N & CH$_4$ \\
 \hline
 {\Large\strut}IRS 46 & $<$1.5 &  $<$52 & $<$0.2 & $<$0.8 & $<$0.6 & $<$3.0 & $<$0.6 & - \\
 GV Tau & $<$0.6 & $<$27 & $<$0.2 & $<$0.9 & $<$0.9 & $<$4.4 & $<$0.9 & - \\
 DR Tau & - & - & - & - & - & - & - & $<$2.3\footnote{From 3 $\mu$m emission data \citep{Mandell2012}.} \\
 \textit{$R=3000$}\footnote{The $R$=3000 and 50,000 values refer to the
   3$\sigma$ limits on the abundance ratios that can be obtained toward
   IRS 46 if observed at higher spectral resolution at a similar S/N.} & 
   \textit{$<$0.9} & \textit{$<$24} & \textit{$<$0.08} & \textit{$<$0.2} & 
   \textit{$<$0.4} & \textit{$<$1.5} & \textit{$<$0.2} & - \\
 \textit{$R=50,000^b$} & \textit{$<$0.3} & \textit{$<$14} & \textit{$<$0.04} & 
   \textit{$<$0.03} & \textit{$<$0.04} & \textit{$<$0.9} & \textit{$<$0.1} & - \\
 \citet{Markwick2002} 1 AU  & - & - & - & 1.0 & 1.0 & 1.0 & 1.0 & 1.1 \\
 \citet{Markwick2002} 5 AU & - & - & - & - & 1.0 & 1.0 & 1.0 & 1.1 \\
 \citet{Najita2011} ref& - & - & - & 4.0E-4 & - & - & - & 0.09 \\
 \citet{Najita2011} O/C = 1.0 & - & - & - & 2.5E-6 & - & - & - & 3.2E-5 \\
 \citet{Agundez2008} 1 AU & - & - & - & - & 0.03 & - & - & 0.01 \\
 \citet{Agundez2008} 3 AU & - & - & - & - & 0.03 & - & - & 0.2 \\
 \citet{Mumma2011} & - & 0.2--20 & - & - & 0.01--0.5 & - & 0.01--0.7 & 0.2--15 \\
 \hline
 \hline
 {\Large\strut}Ratio rel to C$_2$H$_2$ & HCN & C$_4$H$_2$ & CH$_3$CN & SO$_2$ & NH$_3$ & CO$_2$ \\
\hline
 {\Large\strut}IRS 46 & 1.8 & $<$0.3 & $<$370 & $<$11.9 & $<$1.9 & 4.0 \\
 GV Tau & 1.3 & $<$0.2 & $<$190 & $<$13.6 & $<$1.4 & 3.6 \\
 DR Tau & - & - & - & - & - & - \\
 \textit{$R=3000^b$} & - & \textit{$<$0.1} & \textit{$<$150} & \textit{$<$4.0} & \textit{$<$0.7} & - \\
 \textit{$R=50,000^b$} & - & \textit{$<$0.1} & \textit{$<$22} & \textit{$<$1.2} & \textit{$<$0.1} & - \\
 \citet{Markwick2002} 1 AU & 1.0 & 1.0 & 1.0 & 0.9 & 1.0 & 1.0 \\
 \citet{Markwick2002}  5 AU & 1.0 & 1.0 & 1.0 & 0.9 & 1.1 & 1.1 \\
 \citet{Najita2011} ref & 70 & - & - & 8700 & 310 & 110 \\	
 \citet{Najita2011} O/C = 1.0 & 0.9 & - & - & 1.1 & 0.01 & 0.06 \\
 \citet{Agundez2008} 1 AU & 3.0 & - & - & - & 2.0E-4 & 20 \\
 \citet{Agundez2008} 3 AU & 1.0 & - & - & - & 0.03 & 2000 \\
 \citet{Mumma2011} & 0.2--5.0 & - & 0.02--0.4 & 0.2--1.0 & 0.6--20 & 5.0--300 \\ 
 \hline
 \end{tabular} 
\end{minipage}
\label{tab:up_limit_c2h2}
\end{table*}

  \begin{table*}
\caption{Observed molecular column density ratios relative to HCN as well as
CH$_3$CN/NH$_3$ compared with chemical models and with cometary observations.} 
 \begin{minipage}[t]{\columnwidth}
 \renewcommand{\footnoterule}{}
 \centering
  \thispagestyle{empty}
  \begin{tabular}{l l l l l l}
  \hline
   \hline
{\Large\strut}Obs/models & HNC & HC$_3$N & CH$_3$CN & NH$_3$ & CH$_3$CN/NH$_3$ \\
 \hline
 {\Large\strut}IRS 46  & $<$0.3 & $<$0.3 & $<$210 & $<$1.1 & $<$200 \\
 GV Tau  & $<$0.7 & $<$0.7 & $<$140 & $<$1.1 & $<$140 \\
 AS 205 & - & - & - & $<$2.7\footnote{From 3 $\mu$m emission \citep{Mandell2012}.} & - \\
 \textit{$R=3000$}\footnote{The $R$=3000 and 50,000 values refer to the 3$\sigma$ 
   limits on the abundance ratios that can be obtained toward IRS 46 if observed at
   higher spectral resolution at a similar S/N.} & 
   \textit{$<$0.2} & \textit{$<$0.1} & \textit{$<$84} & \textit{$<$0.4} & - \\
 \textit{$R=50,000^b$} & \textit{$<$0.03} & \textit{$<$0.08} & \textit{$<$12} & \textit{$<$0.08} & - \\
 \citet{Markwick2002} at 1 AU &1.0 & 1.0 & 1.0 & 1.0 & 1.0 \\
 \citet{Markwick2002} at 5 AU  & 1.0 & 1.0 & 1.0 & 1.1 & 0.9 \\
 \citet{Najita2011} ref & - & - & 4.0 & - & - \\
 \citet{Najita2011} O/C = 1.0 & - & - & 0.01 & - & - \\
 \citet{Agundez2008} 1 AU & 0.01 & - & - & 6.0e-5 & - \\
 \citet{Agundez2008} 3 AU & 0.03 & - & - & 0.03 & - \\
 \citet{Mumma2011}  & 0.01--0.6 & 0.01--0.9  & 0.02--0.5 & 0.6--25 & 0.01--0.1\\
 \citet{Garrod2008} M & - & - & - & 890 & 7.5E-5 \\
 \hline
 \end{tabular}
 \end{minipage}
\label{tab:up_limit_hcn}
\end{table*}


\section{Discussion} \label{discussion}


Tables~\ref{tab:up_limit_c2h2} and \ref{tab:up_limit_hcn} and Figures~
\ref{fig:comet_disk_c2h2} and \ref{fig:comet_disk_hcn} compare our
limits with a variety of chemical models.  There are two distinct
routes towards molecular complexity in regions of star- and planet
formation. First, at elevated temperatures such as found in the inner
disks, various reactions with activation barriers open up. If atomic
carbon can be liberated from CO and atomic nitrogen from N$_2$, high
abundances of CH$_4$, C$_2$H$_2$ and HCN can be produced. The second
route starts in the pre-stellar cores where ices are formed through
grain surface reactions. At a later stage, these ices can be
transported into the disk and evaporate so that a chemistry rich in
hydrogenated molecules can ensue. We review each of these classes of
models and then discuss our observations in the light of these models
and in comparison with cometary and other data.


\subsection{Warm chemistry} \label{warm}

 \begin{figure*}
\centering
{
 \includegraphics[width=150mm, angle=0.0]{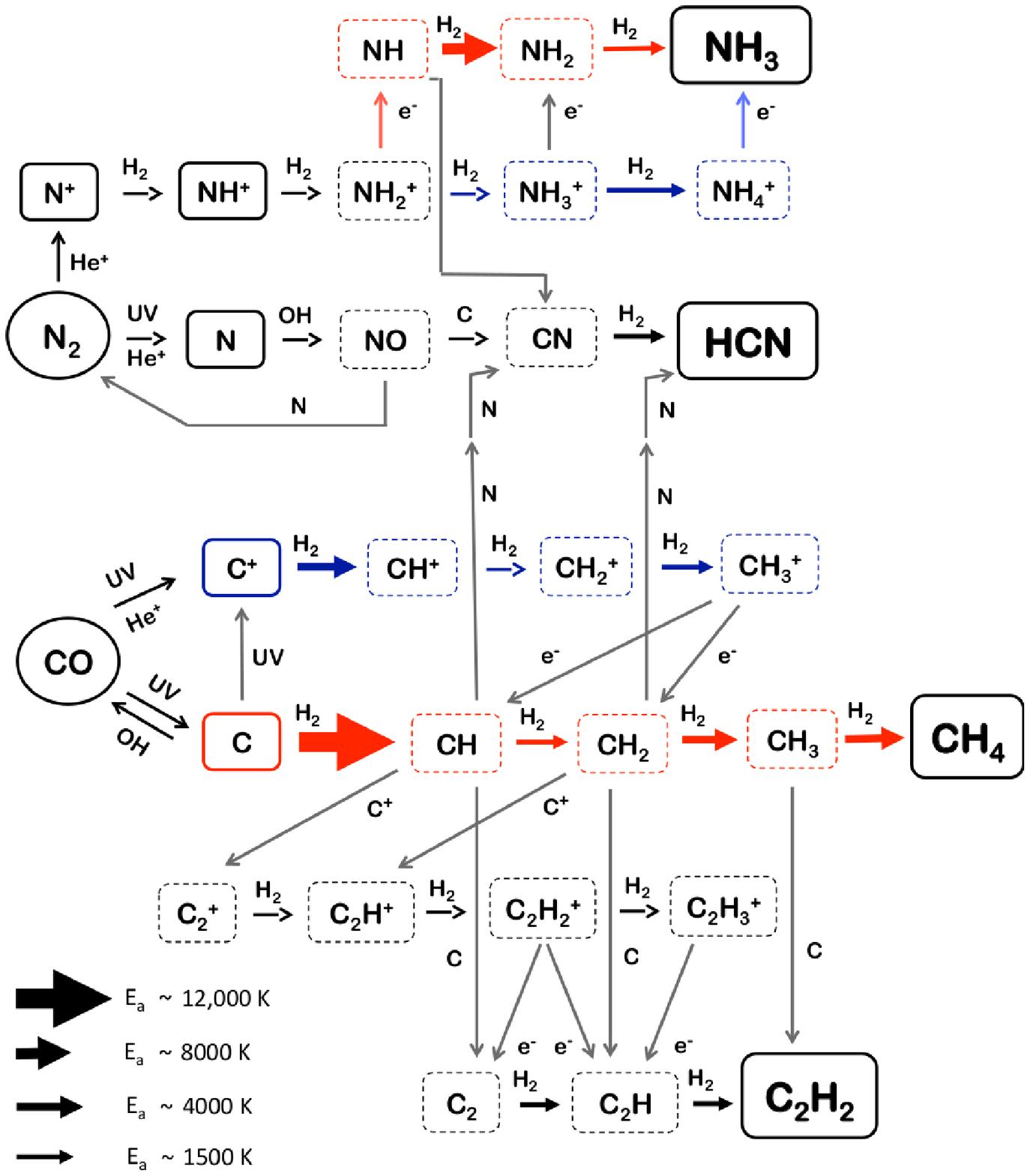}
 \caption{Chemical diagram describing the main reaction routes to form
   HCN, C$_2$H$_2$, CH$_4$ and NH$_3$ in warm gas in protoplanetary
   disks or hot cores depending on the radiation fields, temperature
   and different molecular abundances, adapted from
   \citet{Agundez2008}. At high temperatures of several hundred K the
   reaction routes (in red) starting with C reacting with H$_2$ to CH
   and NH to NH$_2$ dominate over lower temperature chemical routes
   (in blue) of C$^+$ going to CH$^+$ and NH$_3^+$ and NH$_4^+$. The
   different activation energy ($E_a$) barriers for the chemical
   reactions are represented by different types of arrows; the thicker
   the arrow, the higher the activation energy.  }
 \label{fig:chemical_scheme}
}
\end{figure*} 

The warm gas chemistry in the photospheres of disks follows a similar
chemical scheme as that in other interstellar regions with high
temperature gas such as the inner envelopes of massive protostars
\citep[e.g.,][]{Lahuis2000,Doty2002,Rodgers2003,Stauber2005}, high density
photodissociation regions (PDRs) \citep[e.g.,][]{Sternberg1995} and
shocks \citep[e.g.,][]{Mitchell1984,Pineau1987,Viti2011}.  
The chemical scheme starts with
separating C or C$^+$ from CO and N from N$_2$, see
Fig.~\ref{fig:chemical_scheme} \citep[adapted from][]{Agundez2008}.
This can be done either by UV photons, cosmic rays or X-rays.  A
UV-dominated region will produce comparable amounts of C$^+$ and C,
whereas cosmic rays and X-rays produce He$^+$ which will react with CO
to produce primarily C$^+$.

For carbon-bearing species, the rate limiting steps of this scheme are
the reactions of C and C$^+$ with H$_2$ which have activation barriers
$E_a$ of $\sim$12,000 and 4,000 K, respectively.  At temperatures of a
few hundred K, the C$^+$ channel leads to CH$_4$ and C$_2$H$_2$ while
at high temperatures ($>$800 K), the C-channel becomes
active. Reactions of C$_2$ with H$_2$ have activation barriers of
$\sim$1500~K and form another route to produce C$_2$H and subsequently
C$_2$H$_2$.  For nitrogen-bearing species, the reactions of NH with
H$_2$ ($E_a\sim 7800$ K) to form NH$_3$ and of CN with H$_2$ ($E_a
\sim 820$ K) to form HCN also require high temperatures.  At low
temperatures $<200$~K, the above reactions are closed and C$^+$, C and
N will be driven back to CO and N$_2$ through reactions involving
OH. Hence, the higher the temperature the more C$_2$H$_2$, CH$_4$ and
HCN will be produced.  Since the temperature decreases with
  disk radius, the abundances of C$_2$H$_2$, CH$_4$ and HCN show a
  strong radial dependence with a steep decrease beyond 1 AU
  \citep{Agundez2008}.

Similar arguments apply to the more complex hydrocarbons studied
here. In fact, high temperature, high density chemistry starting with
high abundances of C$_2$H$_2$ and HCN resembles the chemistry of the
atmospheres of carbon-rich evolved stars, which has been studied for
decades \citep[e.g.,][]{Cherchneff1993,Millar1994}.  For example,
C$_2$H$_2$ reacts with CH$_3$ produce C$_3$H$_4$, with C$_2$H to
C$_4$H$_2$, and with NH$_3$ to CH$_3$CN. If C$_2$H is sufficiently
abundant (due to photodissociation of C$_2$H$_2$, for example),
subsequent reactions may lead to large unsaturated carbon chains. More
saturated molecules such as C$_2$H$_4$ and C$_2$H$_6$ need CH$_4$ as
their starting point. Finally, nitrogen-containing species are
produced by reactions of HCN or N with hydrocarbons. For example,
C$_2$H$_2$ and C$_2$H react with HCN to form HC$_3$N. It is now well
established that many of these neutral-neutral reactions actually have
substantial rates even at low temperatures ($<$100 K)
\citep{Smith2011}. Thus, the bottleneck in producing the more complex
organic molecules is the formation of the chemical precursors
C$_2$H$_2$, HCN and CH$_4$. The more complex molecules are
  thus also expected to show a strong dependence on disk radius.

Benzene is the most complex molecule probed here and is a precursor
for building larger PAHs.  One route to form benzene is through the
reaction of C$_3$H$_4$ and its ion, found to be important in the inner
disk chemistry in some models \citep{Woods2007}. Alternatively, the
destruction of PAHs by high temperature gas-phase reactions inside the
`soot line' can lead to the production of small hydrocarbons,
including C$_2$H$_2$ and perhaps C$_6$H$_6$ \citep{Kress2010}.

CO$_2$ is the product of the reaction of OH with CO. Models show that
all atomic O is driven into H$_2$O at temperatures above $\sim$230 K
if there are no rapid H$_2$O destruction routes
\citep[e.g.,][]{Draine1983,Charnley1997}. The formation of OH also
needs elevated temperatures, but peaks in the 100--200 K range because
of the rapid reactions of OH with H$_2$ to form H$_2$O at higher
temperatures. In warm gas, the CO$_2$ abundance follows that of OH and
also peaks at 100--200~K, with a strong decrease toward higher
temperatures. Thus, in contrast with C$_2$H$_2$ and HCN, the
  abundance of CO$_2$ increases somewhat beyond 1 AU in disk models,
  until the temperature falls below 100 K \citep{Agundez2008}.

Another parameter that can affect the chemistry is the O/C
  ratio in the gas.  There have been suggestions that this ratio
changes with disk radius due to migration of icy
planetesimals containing a large fraction of the oxygen
\citep{Ciesla2006}. Hydrocarbon abundances are very sensitive to the
O/C ratio \citep{Langer1989,Bettens1995,Najita2011}. As expected,
oxygen-bearing species such as H$_2$O and O$_2$ decrease dramatically
in abundance when O/C is less than 1 because all volatile oxygen
(rather than all carbon) is locked up in CO. Pure carbon species such
as C$_2$H$_2$ increase in abundance by orders of magnitude while
nitrogen species only change appreciably when they contain C (e.g.,
HCN). Since both IRS 46 and GV Tau
show large columns of gaseous CO$_2$, the lines of sight through the
disks probed by the mid-infrared data must pass through gas with
O/C$\geq1$.

Finally, the overall gas/dust ratio can play a role 
\citep[e.g.,][]{Najita2011,Bruderer2012}. 
However, comparison of the CO column density with the
extinction measured from the silicate optical depth shows that this
ratio is close to the interstellar value for both IRS 46 and GV Tau,
indicating no significant grain growth and settling along the lines of
sight through these disks \citep{Kruger2011}.

In \S\ref{models:specific} in the Appendix, more detailed
descriptions of a few specific protoplanetary disk models are given,
focussing on the simpler species which can be compared with
  observations.

\subsection{Surface chemistry}
\label{surface}

The more complex organic molecules such as dimethyl ether seen toward
protostars are the product of an active gas-grain interaction
\citep[see][for reviews]{Tielens1997,Herbst2009}.  
Recent models produce these species on the grains rather than in hot
gas-phase chemistry, either through direct grain surface chemistry or
through mild photolysis of simple ice species resulting in radicals
which then react with each other to form more complex species
\citep{Garrod2008, Oberg2009}. These types of grain chemistry --
either direct surface reactions or photolysis -- will not result in
high abundances of C$_2$H$_2$ and HCN because these species will be
readily hydrogenated by H. On grain surfaces, N is quickly converted
to NH$_3$, while in the gas phase, N flows through NO to N$_2$ with a
slight detour to HCN. Grain surface reactions lead to very high
abundances of NH$_3$, especially in relation to HCN. Observations and comparisons with model predictions show that also CH$_4$ formation is very efficient on dust grains \citep{Oberg2008}.

\subsection{Comparison of models with observations}

Table \ref{tab:up_limit_c2h2} and Figures \ref{fig:comet_disk_c2h2}
and \ref{fig:comet_disk_hcn} show that warm chemistry
can explain the observed abundance ratios of HCN, C$_2$H$_2$ and
CO$_2$ as well as most limits. In the \citet{Markwick2002} models, all
abundance ratios are close to unity, but these ratios refer to the
entire disk rather than just the surface layers. It is therefore not
possible to properly test these models. The models of
\citet{Agundez2008} and \citet{Najita2011} provide column densities
for just the warm surface layers probed by the mid-infrared
  data, but tabulate only a limited number of species see
  \S\ref{models:specific} for details.

The inner disk models of \citet{Agundez2008} provide the correct range
of column densities and abundance ratios in the surface layers around
1 AU. At 3 AU, the C$_2$H$_2$ column has dropped dramatically due to
the lower temperature and the model overproduces the CO$_2$/C$_2$H$_2$
ratio by a factor $>$100. The lower excitation temperatures
  observed for CO$_2$ in our data compared with C$_2$H$_2$ and HCN
  provide indirect evidence for the radial dependence of the
  chemistry, with CO$_2$ peaking at larger distances (see \S~\ref{warm}).
The pure XDR model results given by \citet{Najita2011}, for
the reference case of a disk at 1 AU, agree less well with the
observations than the PDR model results at 1 AU. For example the
SO$_2$, NH$_3$ and CO$_2$ model ratios relative to C$_2$H$_2$ are
several orders of magnitude higher than our observed upper limits and
detections.  This can be largely explained by the fact that
  the \citet{Najita2011} models do not contain photodissociation
  reactions in the surface layers which limit their abundances.
Lowering the O/C ratio to 1 instead of 2.5 decreases the difference
between the observations and the pure X-ray models but now
underproduces CO$_2$. As shown by \citet{Walsh2012} in
  combined UV + X-ray models, the inclusion of UV-induced processes is
  generally more important than those due to X-rays.

The importance of gas-grain interaction for the composition of the
inner regions of the disk may well be revealed by NH$_3$ and CH$_4$
searches. As mentioned in section \ref{surface}, high abundances of
CH$_4$ and NH$_3$ relative to HCN and C$_2$H$_2$ could indicate that
the chemistry in the inner part of disks has been reset due to
evaporation of ices from dust grains. Table \ref{tab:up_limit_hcn}
includes the results from model M of \citet{Garrod2008}. The
efficiency to form NH$_3$ on dust grains is the main reason why
NH$_3$/HCN and CH$_3$CN/NH$_3$ are so different in the grain surface
models of \citet{Garrod2008} compared with the warm chemistry models
developed by \citet{Najita2011} and \citet{Agundez2008}.
Currently no observational limit can be set on the
  CH$_3$CN/NH$_3$ ratio.  However, our observed limit on NH$_3$/HCN
clearly favors a low ratio for the abundances of these species,
inconsistent with a significant contribution from pure ice
chemistry.

The observed upper limits on the other species investigated here are
in general higher or comparable to what the chemical models
predict.  Thus, a firm conclusion from our data is that the molecular
abundance ratios cannot be higher than what is predicted in the
current models.  However they can be lower and future higher
  resolution observations with for example JWST or ELTs that go an
  order of magnitude (or more) deeper may be able to directly test the
  chemical models (see Table~\ref{tab:up_limit_c2h2}). Specifically,
  deep searches for HNC, HC$_3$N, C$_6$H$_6$, SO$_2$ and NH$_3$ should
  be able to distinguish between some of the models, especially if the
  $S/N$ ratio of the data can be pushed to values significantly higher
  than 100.

\subsection{Comparison with protostars, other disks and comets}

The abundance ratios of the detected molecules ---HCN, C$_2$H$_2$ and
CO$_2$--- are remarkably close (within factors of two) to those
observed toward high-mass protostars \citep{Lahuis2000}, which have
been interpreted with high temperature gas-phase chemistry models
\citep{Doty2002}.  Since our NH$_3$/HCN limits of $<$1.1 are close to
the detected NH$_3$/HCN ratio of $\sim$1.2 toward one high-mass
protostar \citep{Knez2009}, this suggests that deeper high resolution
searches for NH$_3$ may be fruitful.

It is also interesting to compare the two observed C$_2$H$_2$/HCN
abundance ratios of 0.6 and 0.8 for IRS 46 and GV Tau respectively
with the one order of magnitude lower values presented in
\citet{Carr2011} for 5 protoplanetary disks in Taurus. Those ratios
have been inferred from emission lines, whereas ours come from
absorption data and may probe a different part of the disk 
($<1$ AU vs up to a few AU).
This could imply that the C$_2$H$_2$/HCN abundance ratios probed through
our edge-on disks presented are more similar to the hot cores and
comets in Fig.~12 of \citet{Carr2011} than to their protoplanetary
disks. 
However, \citet{Salyk2011a} and \citet{Mandell2012} present
C$_2$H$_2$/HCN abundance ratios of $\sim$1 for several protoplanetary
disks, illustrating that different
  analysis methods of the emission data can lead to an order of
  magnitude different abundance ratios. The latter values are about
the same as the abundance ratios found in the hot core observations
and in our disk absorption data.

It is interesting to further compare our inferred molecular
ratios to those
observed in comets, see Table \ref{tab:up_limit_c2h2} and Figures
\ref{fig:comet_disk_c2h2} and \ref{fig:comet_disk_hcn}. 
The ratios tabulated for comets are presented as ranges between the
highest and lowest observed ratios \citep{Mumma2011}, and are all
within our observed upper limits and detections.  This comparison
suggests that some of the HCN and C$_2$H$_2$ produced by warm
chemistry in the inner disk may be incorporated into cold
comets. However, the presence of complex organics such as dimethyl
ether in high abundances in comets \citep[see][for a recent
review]{Bockelee2011} suggests that the ice chemistry route is also
important for their organic inventory.  The detection of
  additional species in the inner disk such as CH$_4$, HNC and HC$_3$N
  will further constrain their relative importance.

In summary, our observed ratios suggest that warm chemistry models
with O/C$>$1 are most relevant for explaining the observed
abundance patterns in disks. Future higher resolution observations of
these molecules with JWST-MIRI and other facilities can help in
answering the question to what extent warm chemistry and surface
chemistry contribute to the chemical composition of the gas in the
planet- and comet-forming zones of disks.

\section{Conclusions} \label{conclusions}

In this paper, mid-infrared spectra of HCN, C$_2$H$_2$ and CO$_2$ have
been analyzed for two edge-on disks, IRS 46 and GV Tau. The high $S/N$
data have also been used to put upper limits on the abundances of
other molecules predicted to be abundant in the inner disk. The main
conclusions are:

\begin{itemize}
\item[$\bullet$] The two disks have similar column densities and
  similar abundance ratios of warm HCN, C$_2$H$_2$ and CO$_2$. The
  first two molecules probe gas with excitation temperatures $T \sim
  400-700$~K, whereas CO$_2$ probes somewhat cooler gas. These results
  are similar to those found toward massive protostars. 

\item[$\bullet$] The observed abundance ratios for these three
    molecules are consistent with high temperature inner disk
    chemistry models with O/C$>$1 in which the abundances of HCN and
    C$_2$H$_2$ rapidly drop with radius beyond 1 AU due to the
    decrease in temperature but that of CO$_2$ peaks at larger
    radius.

\item[$\bullet$] No other absorption features are detected above
  3$\sigma$ in either source, providing upper limits on a variety of
  hydrocarbon molecules, NH$_3$ and SO$_2$ that are of order unity or
  less with respect to C$_2$H$_2$ or HCN.

\item[$\bullet$] The upper limits relative to
  C$_2$H$_2$ and HCN are either higher or close to values given by
  high temperature chemistry models of protoplanetary disks.  The
  observed NH$_3$/HCN limit is much lower than would be expected if
  the chemistry in disks would have been reset due to evaporation of
  icy mantles on dust grains.

\item[$\bullet$] Hot chemistry disk models including both UV radiation
  and X-rays produce abundance ratios in better agreement with our
  observations than pure X-ray models.

\item[$\bullet$] The observed abundance ratios in comets are within
  the same range as our observed ratios or upper limits.  The
  composition of comets could therefore be partly build up from gas in
  the inner regions of protoplanetary disks mixed outward to the
  comet-forming zone.

\item[$\bullet$] Future observations using higher resolution
  instruments on JWST, ELT, SPICA and SOFIA will be able to detect
  column densities which are an order of magnitude or more
  lower than the upper limits extracted from the \textit{Spitzer}-IRS
  data, especially if $S/N$ ratios much higher than 100 can be
    obtained.  Such data would provide much better constraints of the
  hot gas phase chemical models of the inner disk.  Edge-on systems
  such as IRS 46 and GV Tau remain uniquely suited for this purpose.

\end{itemize}

  \begin{acknowledgements}
    The authors are grateful to Catherine Walsh for further
    information on inner disk models.  JEB is supported by grant
    614.000.605 from Netherlands Organization of Scientific Research
    (NWO). EvD and FL acknowledge support from a NWO Spinoza Grant,
    from the Netherlands Research School for Astronomy (NOVA) and from
    A-ERC grant 291141-CHEMPLAN.
 \end{acknowledgements}

\clearpage




\Online
\newpage
\newpage

\appendix

\section{Discussion of specific disk models}
\label{models:specific}

In this section, we provide more details of the specific disk models
to which our observational results are compared.

\subsection{X-ray dominated region (XDR) surface layers}
\label{models:xdr}

\citet{Najita2011} have analyzed the chemistry of the inner (0.25--20
AU) portions of protoplanetary disks exposed to X rays. The physical
structure derives from the thermochemical model developed by
\citet{Glassgold2009}. The thermal structure is evaluated separately
for dust and gas and the surface gas temperature significantly exceeds
that of the dust in the upper layers. In their model, gas and dust
temperatures are decoupled for column densities less than 10$^{22}$
cm$^{-2}$.  If only the disk surface is considered ($N<$
3$^.$10$^{21}$ cm$^{-2}$), gas temperatures of 300 K are reached out
to radial distances of 4 AU.

The formation of hydrocarbons in this model is triggered by X-rays
which produce He$^+$ which liberates the C$^+$ from CO and N from N$_2$. Acetylene is
subsequently produced by C and C$^+$ insertion reactions with small hydrocarbon
radicals (Fig.~\ref{fig:chemical_scheme}). Hence, the C$_2$H$_2$
abundance is sensitive to the X-ray ionization rate. Radicals such as
OH are also sensitive to X-rays. At low temperature, OH results from
the recombination of H$_3$O$^+$ produced by ion-molecule reactions and
its abundance scales directly with the ionization rate. The formation
rate of daughter species of OH such as NO, SO, SO$_2$ and CO$_2$ is similarly
increased by X-ray ionization.
However, if their destruction is also dominated by He$^+$, their
abundances are not sensitive to the X-ray ionization flux. This is the
case for CO$_2$, for example. Note that the models of
\citet{Najita2011} do not include UV photodissociation and may thus
underestimate the amount of OH that could be produced from H$_2$O in
the upper layers. 

The transformation of N back to N$_2$ is mediated by neutral reactions and
is accompanied by significant column densities of warm
nitrogen-bearing molecules such as HCN and NH$_3$ in these
models. Specifically, HCN is influenced by X-rays in its
formation route both because it liberates N from N$_2$ and then N
reacts with OH to form NO, with NO subsequently reacting with C to
form CN. HCN is then formed through reactions of CN with
H$_2$. Because of lack of photodissociation in the model, HCN is destroyed by
He$^+$ and hence destruction is also sensitive to the X-ray. In the
end, the HCN decreases slightly with X-ray luminosity in the models by
\citet{Najita2011}.

\subsection{Photodissociation region (PDR) surface layers}
\label{models:pdr}

Various recent models have analyzed the gas phase chemistry of hot
inner regions of protoplanetary disks including UV radiation for the
chemistry and heating of the gas
\citep{Agundez2008,Woitke2009,Willacy2009,Vasyunin2011,Walsh2012}. The
\citet{Agundez2008} study is particularly instructive because they
present models for just the photosphere of the disk, down to
H$_2$ column densities of $5\times 10^{21}$ cm$^{-2}$ to which the UV
penetrates.
In their models, FUV photons and cosmic rays produce C, C$^+$ and
N. Atomic N is then channeled to HCN through a similar reaction
routine as that described in Fig.~\ref{fig:chemical_scheme}. The C$^+$
leads to a rich hydrocarbon radical chemistry. However, the formation
of high abundances of C$_2$H$_2$ and CH$_4$ requires the reaction of
atomic C with H$_2$ to proceed which has a very high activation
barrier. This reaction only proceeds in very warm gas ($T >$
500~K) forming CH. 
This is reflected in the radial dependence of the C$_2$H$_2$ and
CH$_4$ columns, which reach values of $\sim$10$^{16}$ cm$^{-2}$ out to
radii of $\sim$1 AU, but then drop by orders of magnitude in the
colder gas. This rapid drop with disk radius is also seen in models by
other authors. In contrast, the CO$_2$ column increases with radius in
the inner 1 AU since it favors somewhat colder gas.

\citet{Walsh2012} present a combined UV + X-ray model which shows that
a correct treatment of the photodissociation is generally more
important than including X-rays.  For the important species considered
here ---C$_2$H$_2$, HCN, CO$_2$, CH$_4$ and NH$_3$--- the column
densities in the inner disk do not change measurably when X-rays
are added to the UV model.

\section{Auxilliary figures}
\label{figures:appendix}

This appendix presents simulations of the spectra of all molecules
considered here at higher spectral resolving power of $R=3000$ and
$R=50,000$, appropriate for future instruments. In addition, spectra at
the \textit{Spitzer} resolving power of $R=600$ are included.  All spectra
are computed for $T_{\rm ex}=200$, 500 and 1000\,K, $b=5$ km s$^{-1}$
and a column density of 1$^.$10$^{16}$ cm$^{-2}$.

\clearpage

 \begin{figure*}
\centering
{
 \includegraphics[width=160mm, angle=0.0]{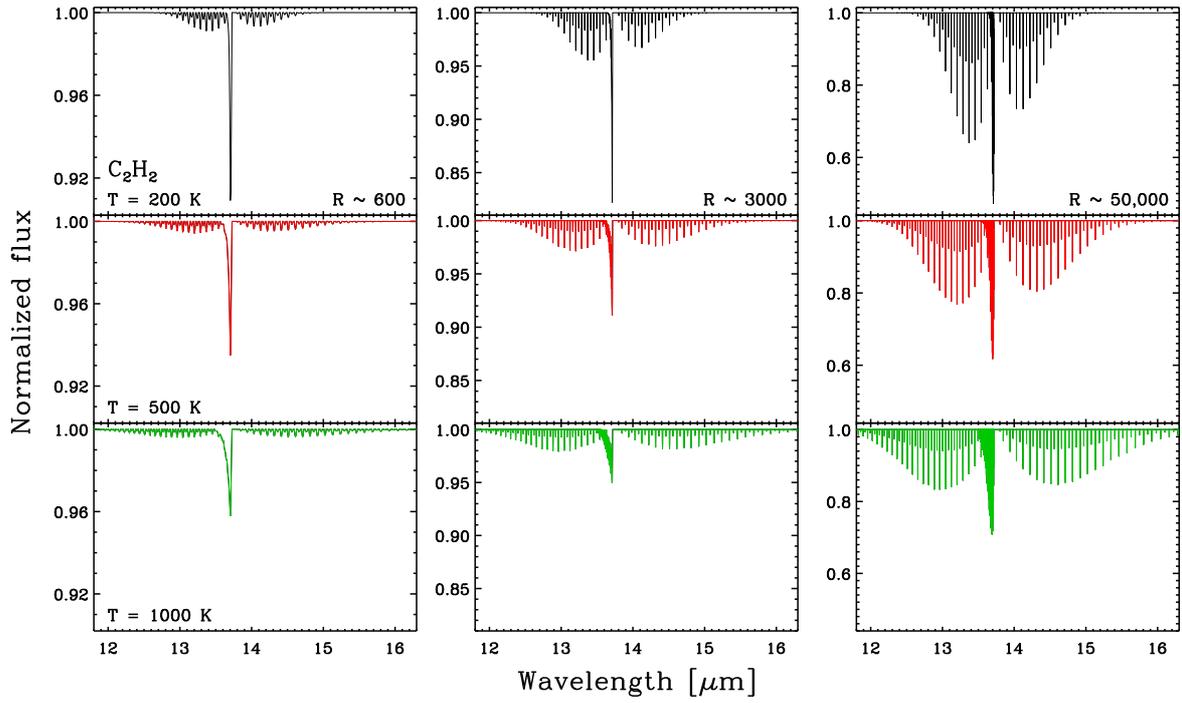}
 \caption{The synthetic spectrum of C$_2$H$_2$ at a column density of
   1.0$^.$10$^{16}$ cm$^{-2}$, excitation temperatures of 200 (top),
   500 (middle) and 1000 K (bottom), and spectral resolving powers of
   600 (left), 3000 (middle) and 50,000 (right). Note the different
   vertical scales for the different spectral resolving powers in this
   and subsequent figures.}
 \label{fig:c2h2_plot_tot}
}
\end{figure*}

 \begin{figure*}
\centering
{
 \includegraphics[width=160mm, angle=0.0]{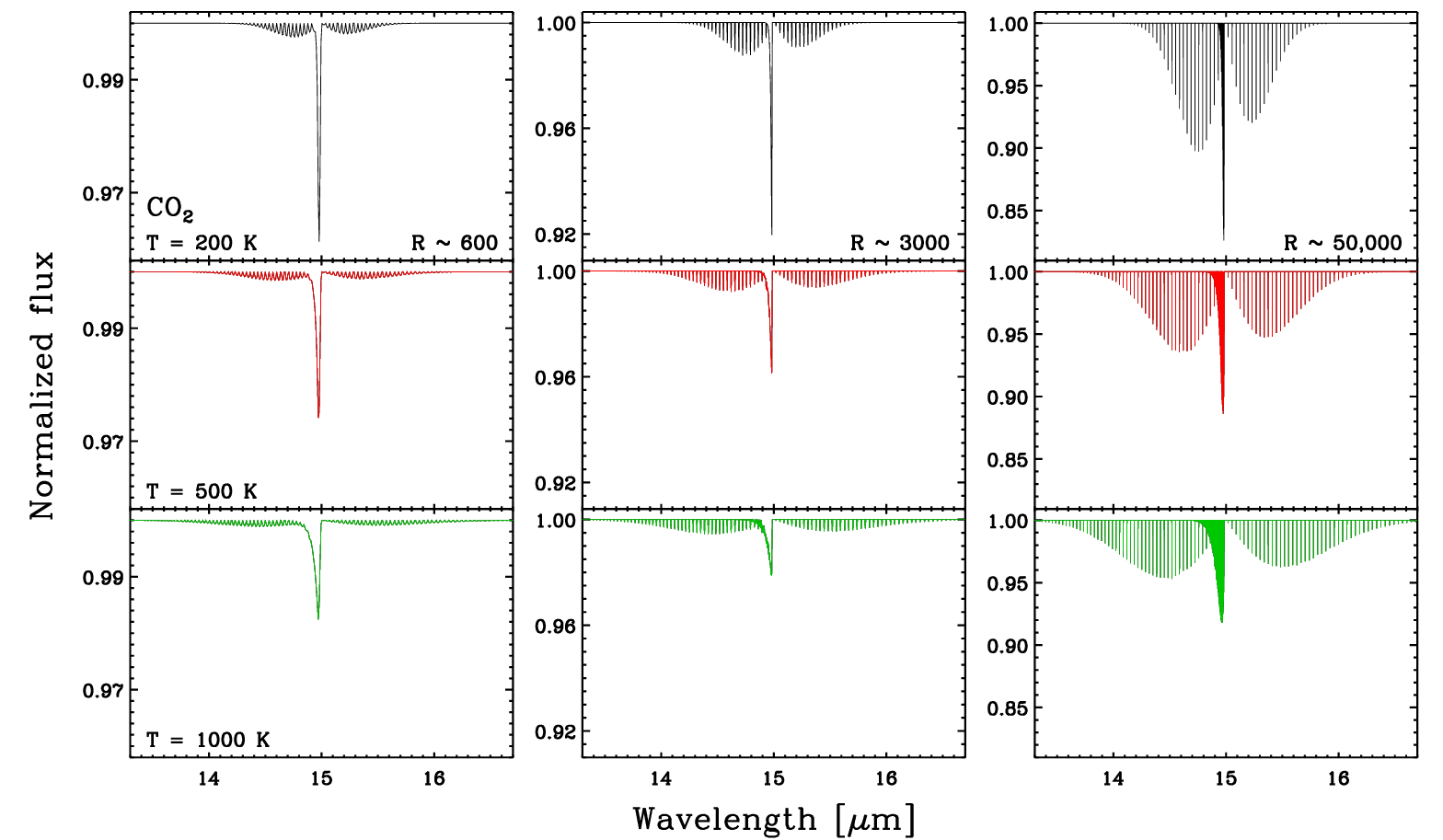}
 \caption{The synthetic spectrum of CO$_2$ at a column density of 1.0$^.$10$^{16}$ cm$^{-2}$, excitation temperatures of 200 (top),
   500 (middle) and 1000 K (bottom), and spectral resolving powers of
   600 (left), 3000 (middle) and 50,000 (right).}
 \label{fig:co2_plot_tot}
}
\end{figure*} 

\clearpage

 \begin{figure*}
\centering
{
 \includegraphics[width=160mm, angle=0.0]{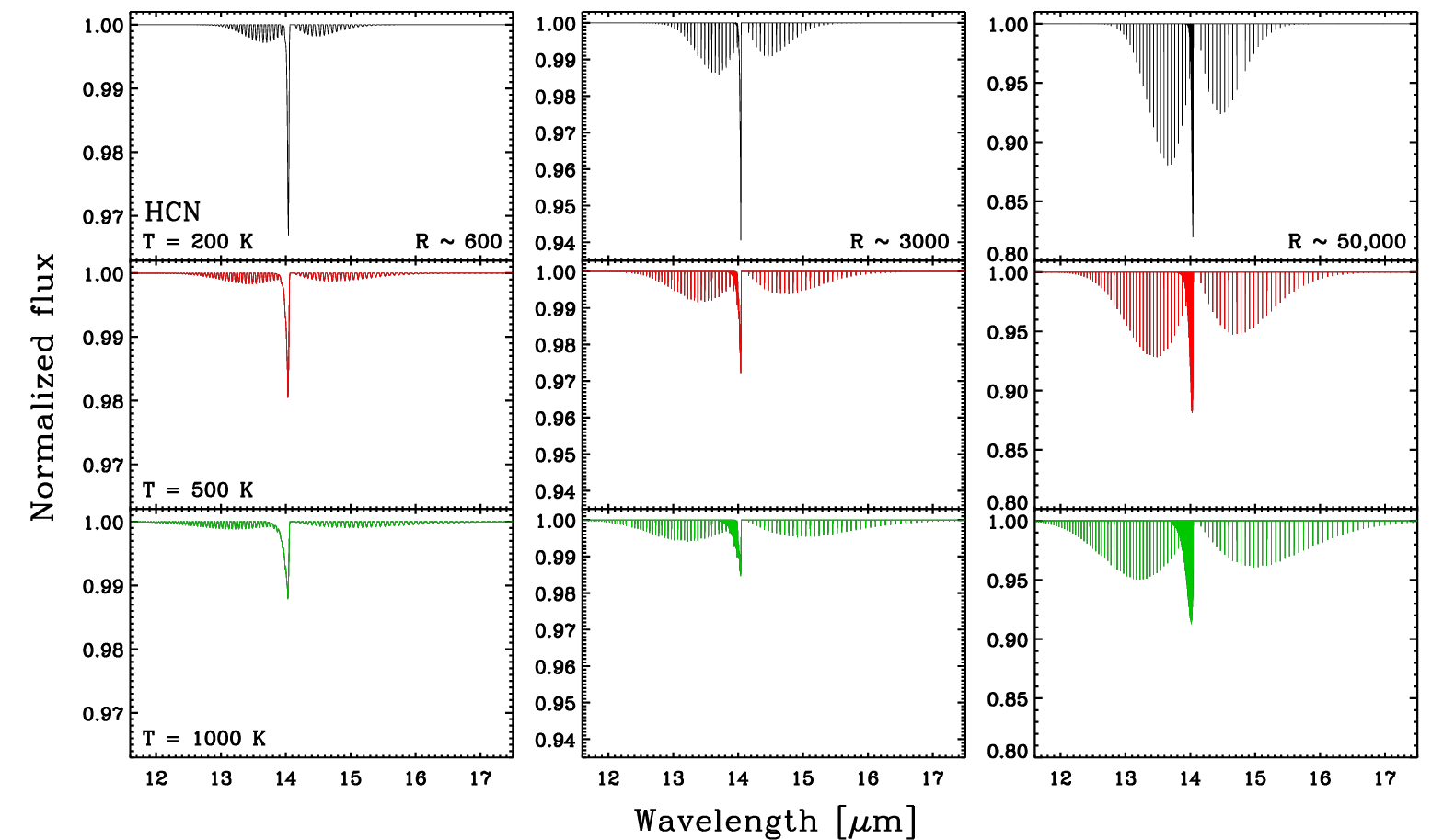}
 \caption{The synthetic spectrum of HCN at a column density of 1.0$^.$10$^{16}$ cm$^{-2}$, 
excitation temperatures of 200 (top),
   500 (middle) and 1000 K (bottom), and spectral resolving powers of
   600 (left), 3000 (middle) and 50,000 (right).}
 \label{fig:hcn_plot_tot}
}
\end{figure*}

 \begin{figure*}
\centering
{
 \includegraphics[width=160mm, angle=0.0]{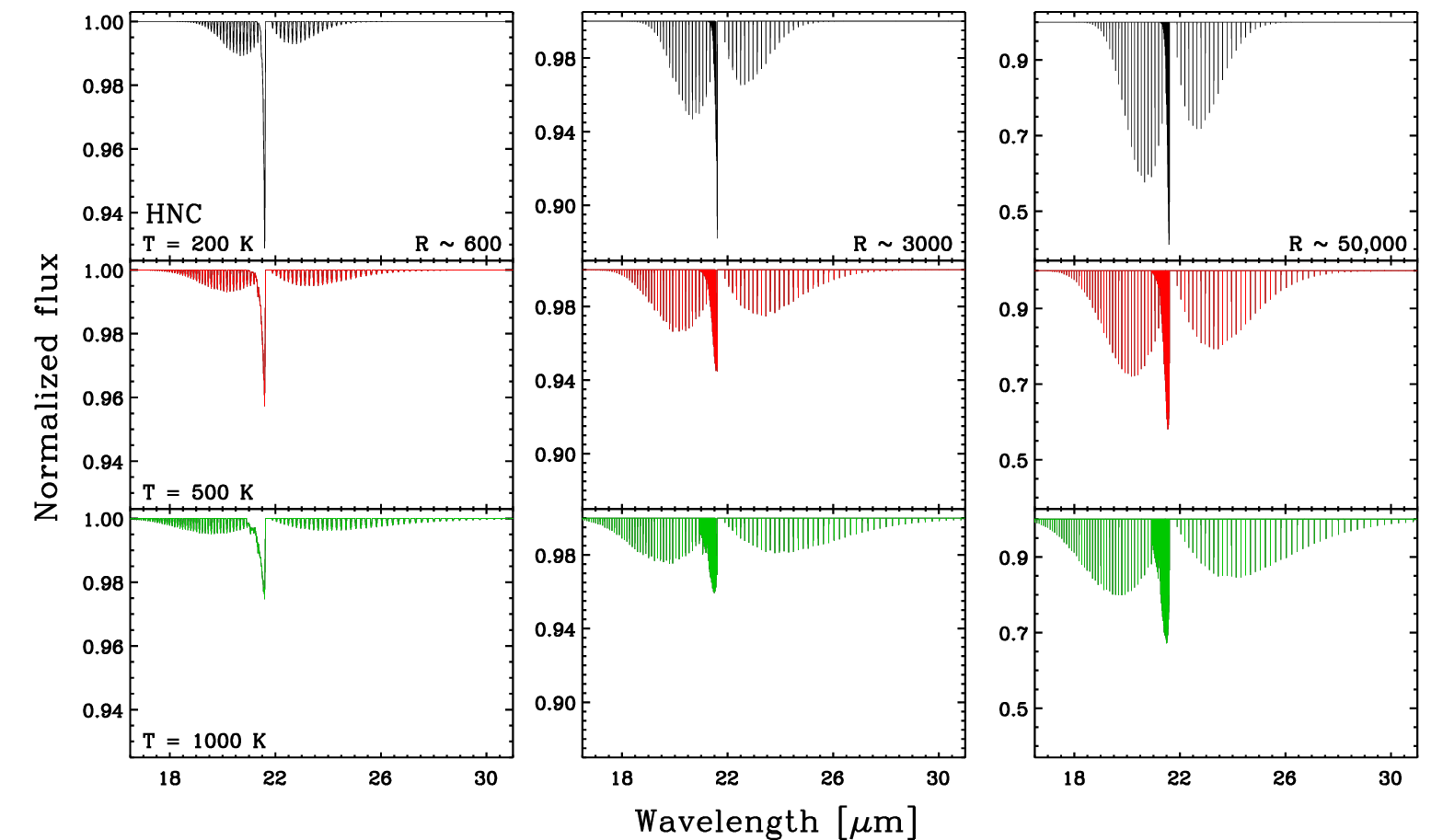}
 \caption{The synthetic spectrum of HNC at a column density of 1.0$^.$10$^{16}$ cm$^{-2}$, 
excitation temperatures of 200 (top),
   500 (middle) and 1000 K (bottom), and spectral resolving powers of
   600 (left), 3000 (middle) and 50,000 (right).}
 \label{fig:hnc_plot_tot}
}
\end{figure*} 

\clearpage

 \begin{figure*}
\centering
{
 \includegraphics[width=160mm, angle=0.0]{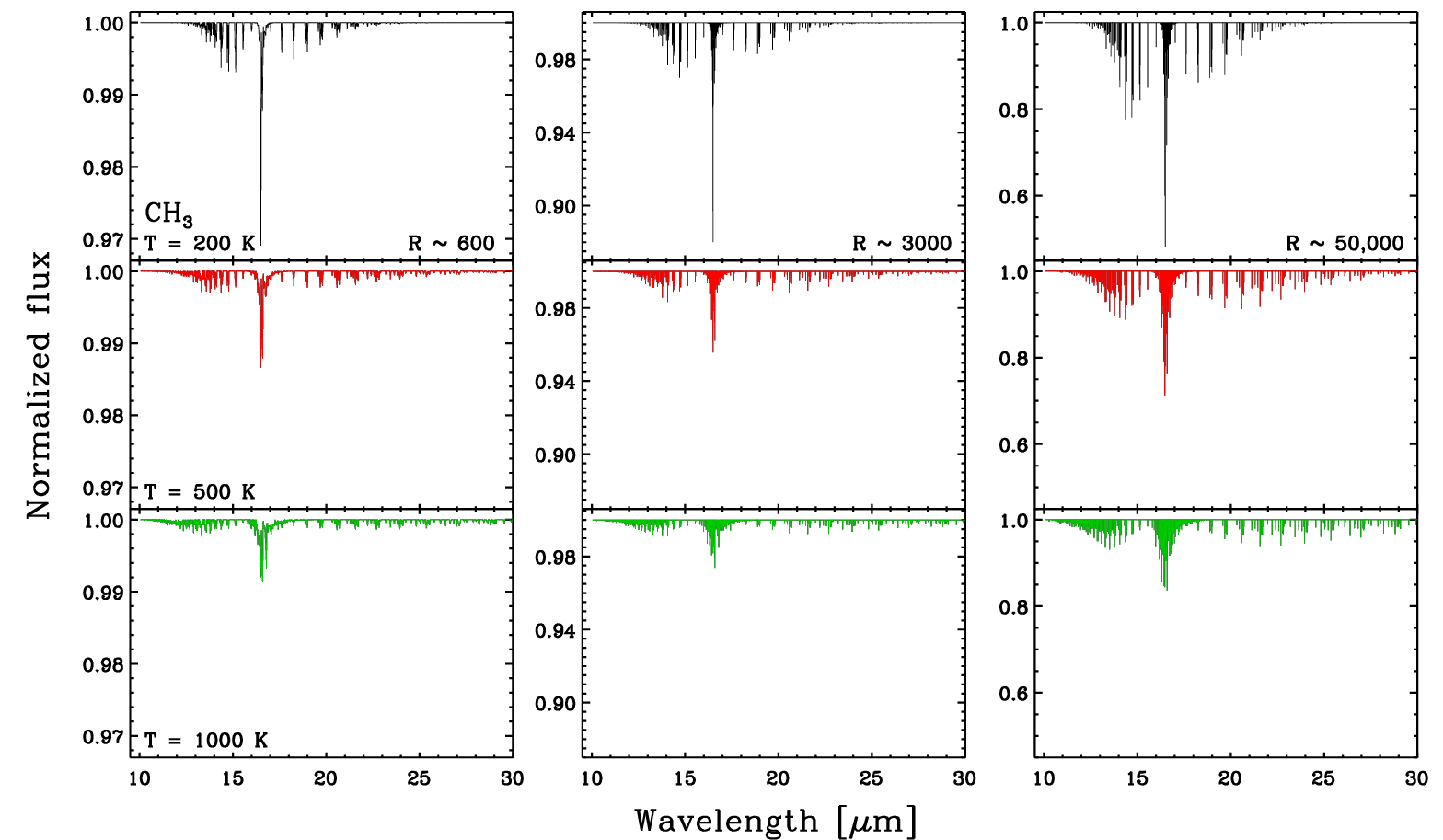}
 \caption{The synthetic spectrum of CH$_3$ at a column density of 1.0$^.$10$^{16}$ cm$^{-2}$, 
excitation temperatures of 200 (top),
   500 (middle) and 1000 K (bottom), and spectral resolving powers of
   600 (left), 3000 (middle) and 50,000 (right).
}
 \label{fig:ch3_plot_tot}
}
\end{figure*} 

 \begin{figure*}
\centering
{
 \includegraphics[width=160mm, angle=0.0]{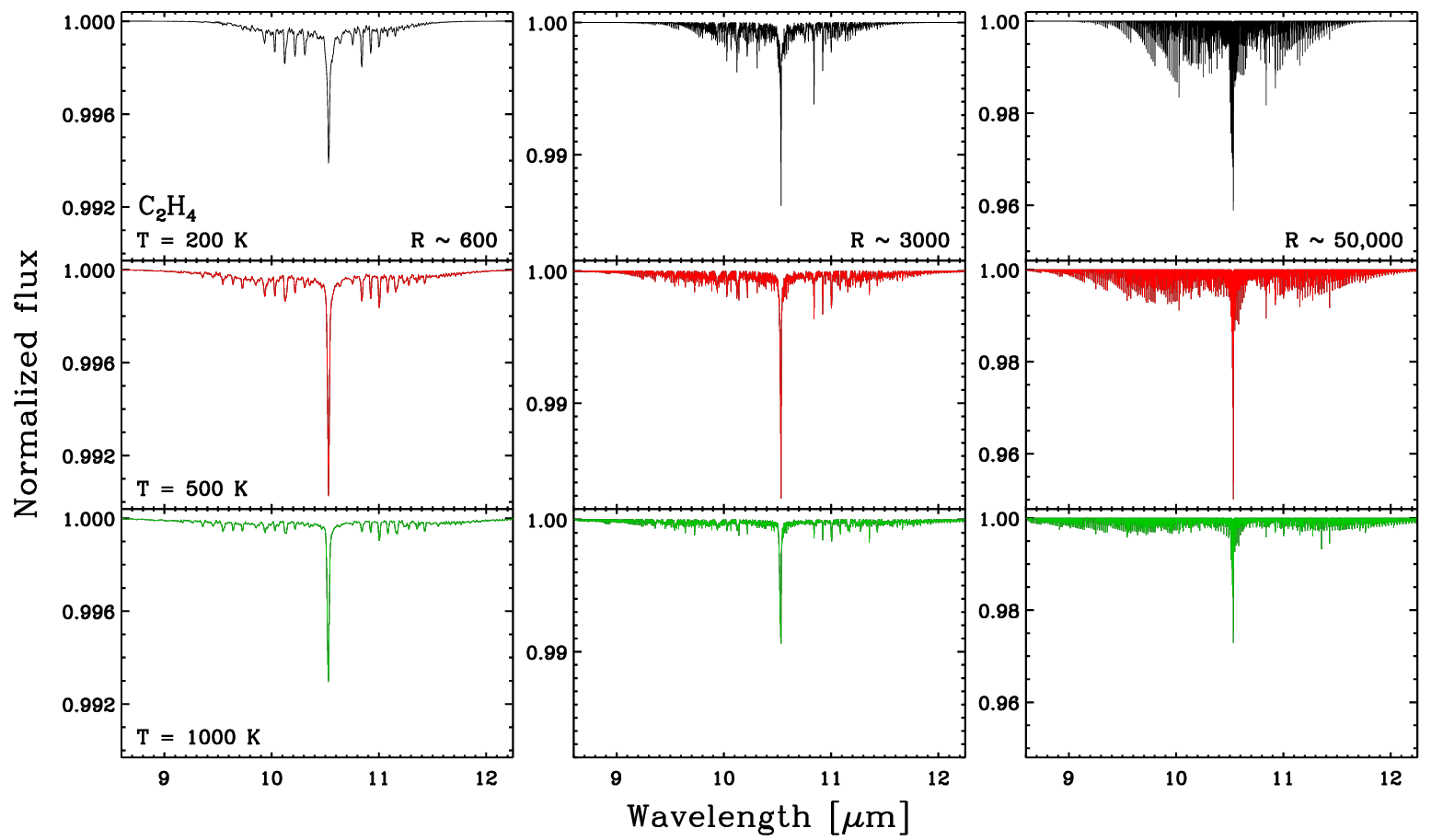}
 \caption{The synthetic spectrum of C$_2$H$_4$ at a column density of 1.0$^.$10$^{16}$ cm$^{-2}$, 
excitation temperatures of 200 (top),
   500 (middle) and 1000 K (bottom), and spectral resolving powers of
   600 (left), 3000 (middle) and 50,000 (right).}
 \label{fig:c2h4_plot_tot}
}
\end{figure*} 

\clearpage


 \begin{figure*}
 \centering
{
 \includegraphics[width=160mm, angle=0.0]{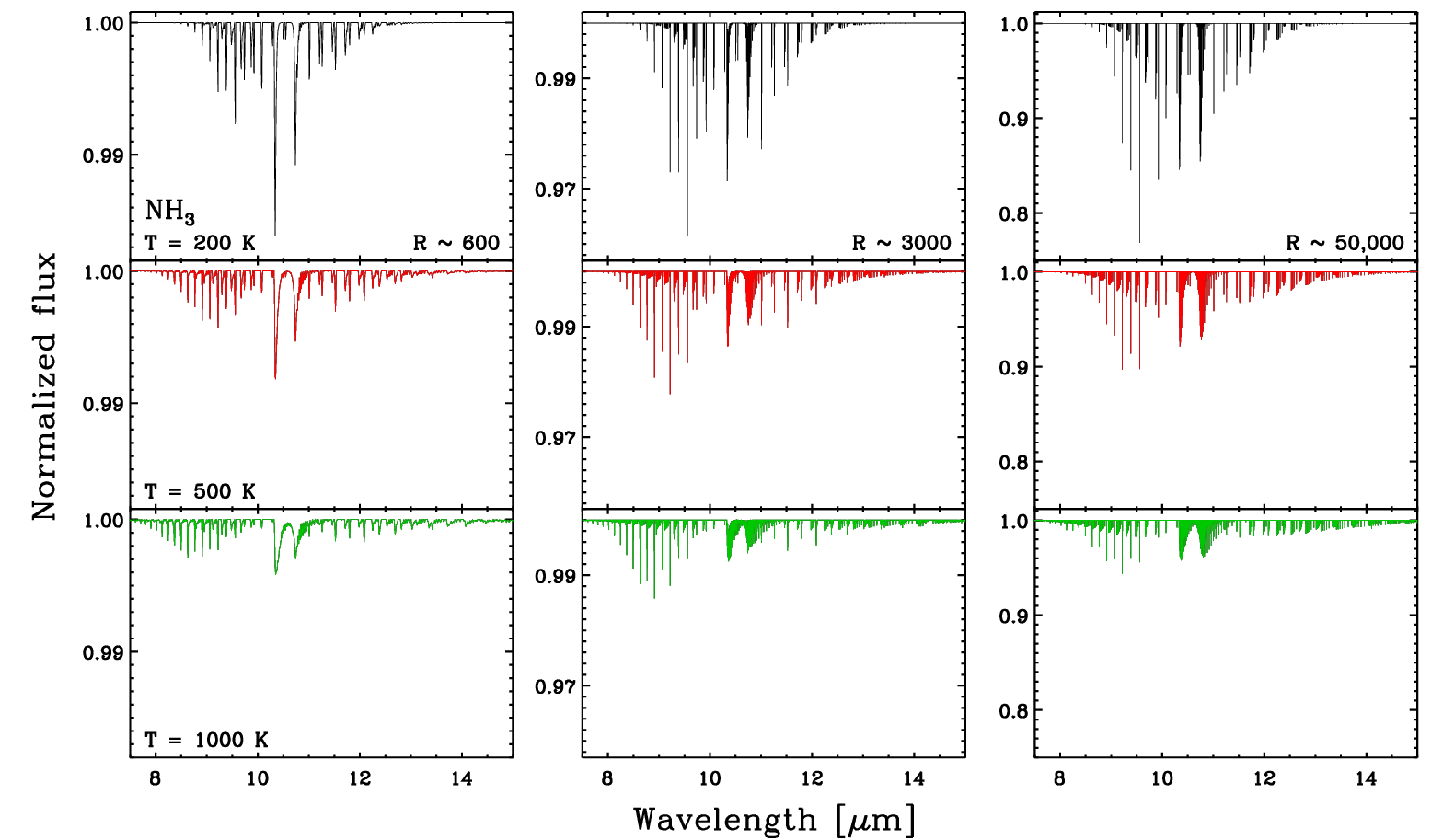}
 \caption{The synthetic spectrum of NH$_3$ at a column density of 1.0$^.$10$^{16}$ cm$^{-2}$, 
excitation temperatures of 200 (top),
   500 (middle) and 1000 K (bottom), and spectral resolving powers of
   600 (left), 3000 (middle) and 50,000 (right).}
 \label{fig:nh3_plot_tot}
}
\end{figure*} 

 \begin{figure*}
\centering
{
 \includegraphics[width=160mm, angle=0.0]{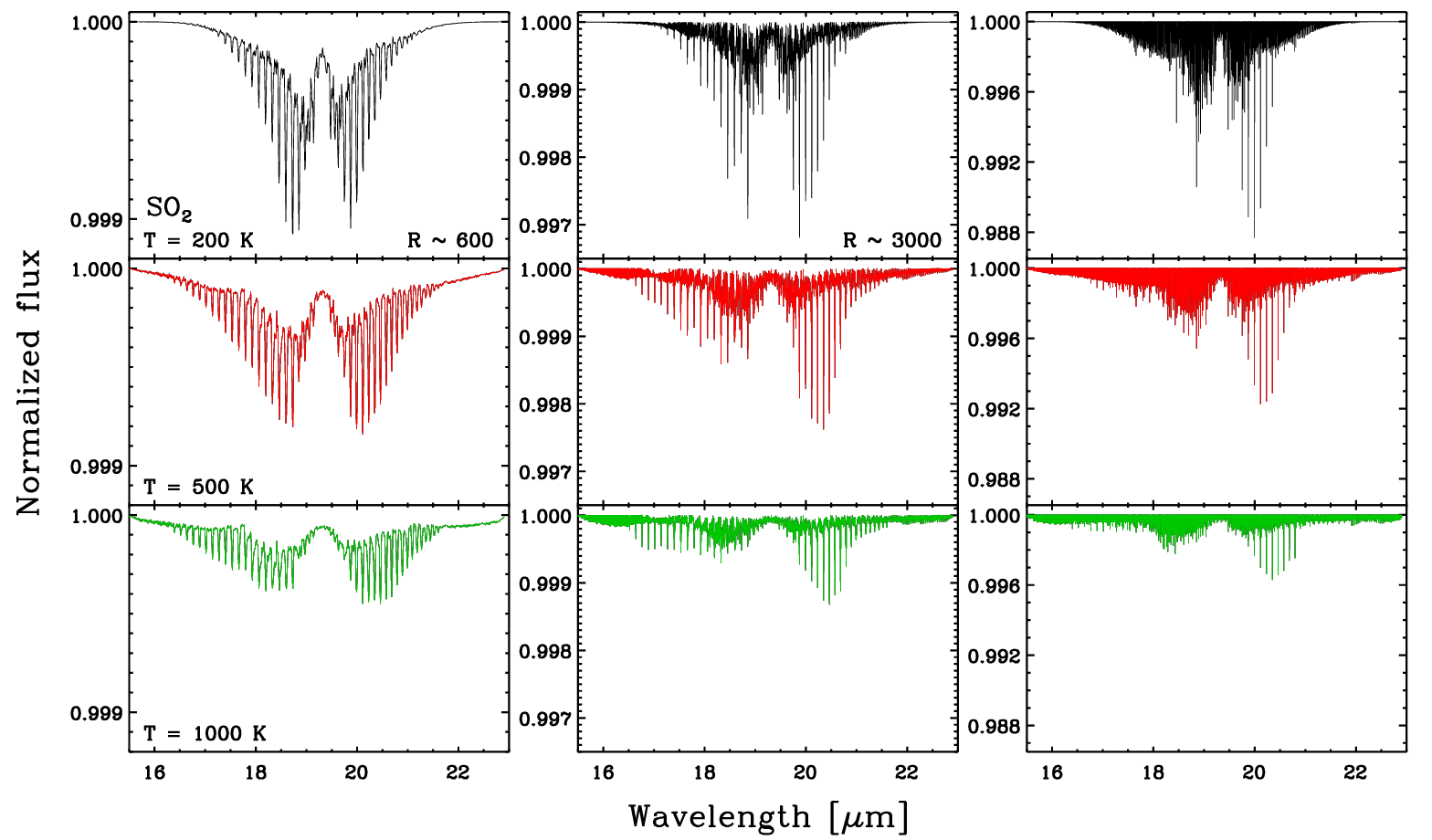}
 \caption{The synthetic spectrum of SO$_2$ at a column density of 1.0$^.$10$^{16}$ cm$^{-2}$, 
excitation temperatures of 200 (top),
   500 (middle) and 1000 K (bottom), and spectral resolving powers of
   600 (left), 3000 (middle) and 50,000 (right).
}
 \label{fig:so2_plot_tot}
}
\end{figure*} 

\clearpage

 \begin{figure*}
\centering
{
 \includegraphics[width=160mm, angle=0.0]{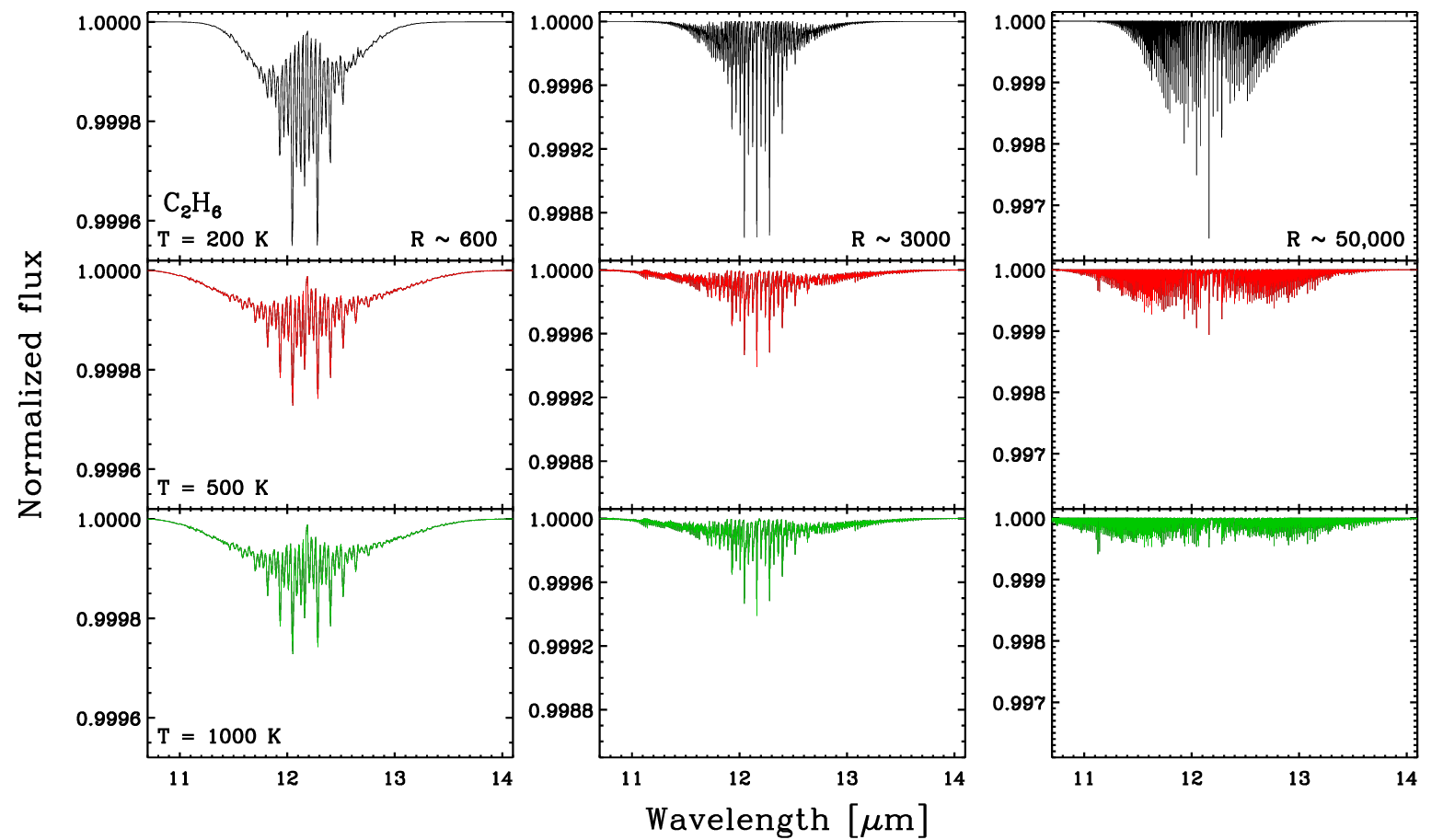}
 \caption{The synthetic spectrum of C$_2$H$_6$ at a column density of 1.0$^.$10$^{16}$ cm$^{-2}$, 
excitation temperatures of 200 (top),
   500 (middle) and 1000 K (bottom), and spectral resolving powers of
   600 (left), 3000 (middle) and 50,000 (right).
}
 \label{fig:c2h6_plot_tot}
}
\end{figure*} 

 \begin{figure*}
\centering
{
 \includegraphics[width=160mm, angle=0.0]{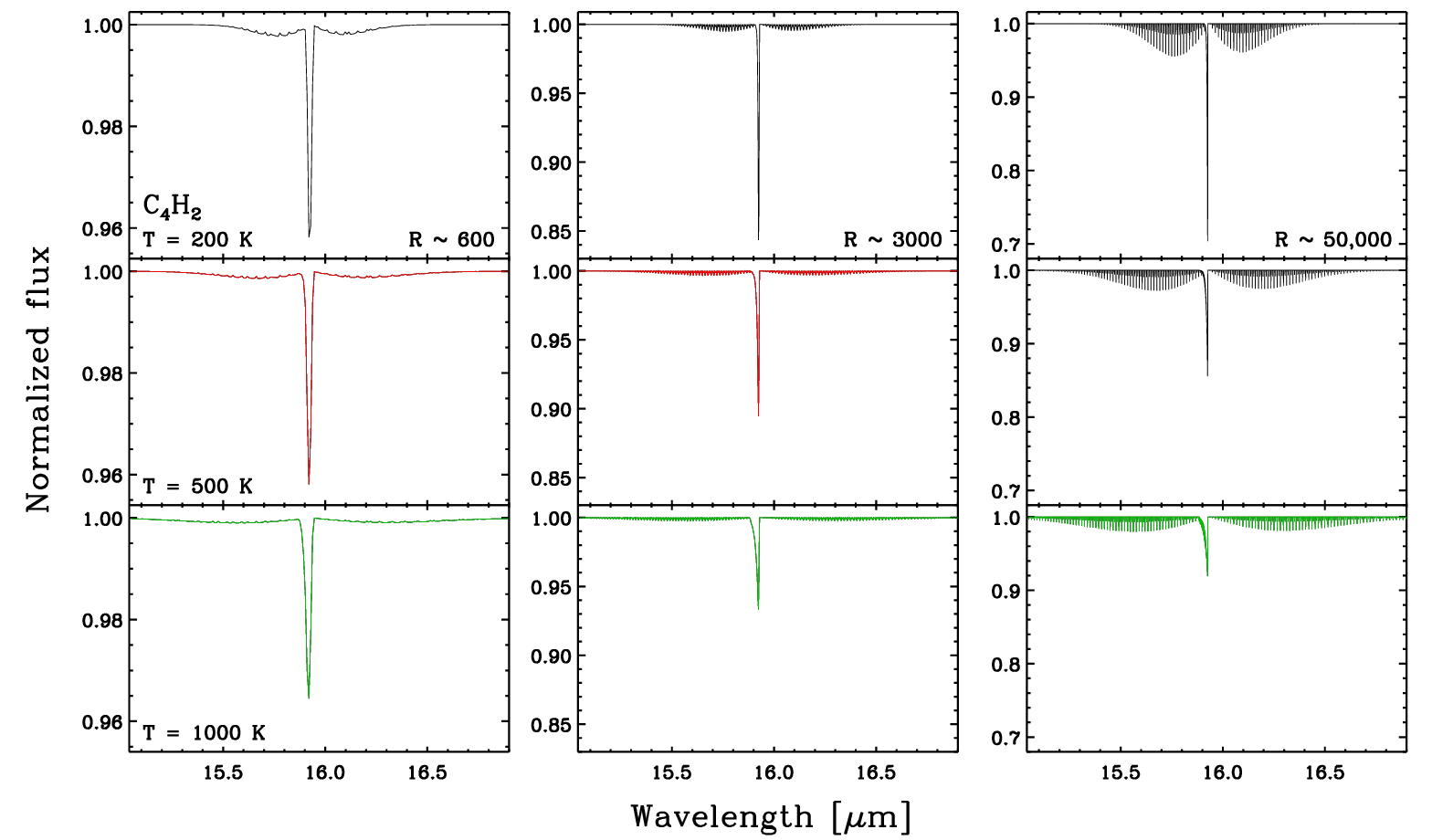}
 \caption{The synthetic spectrum of C$_4$H$_2$ at a column density of 1.0$^.$10$^{16}$ cm$^{-2}$, excitation temperatures of 200 (top),
   500 (middle) and 1000 K (bottom), and spectral resolving powers of
   600 (left), 3000 (middle) and 50,000 (right).
}
 \label{fig:c4h2_plot_tot}
}
\end{figure*} 

\clearpage

\begin{figure*}
\centering
{
 \includegraphics[width=160mm, angle=0.0]{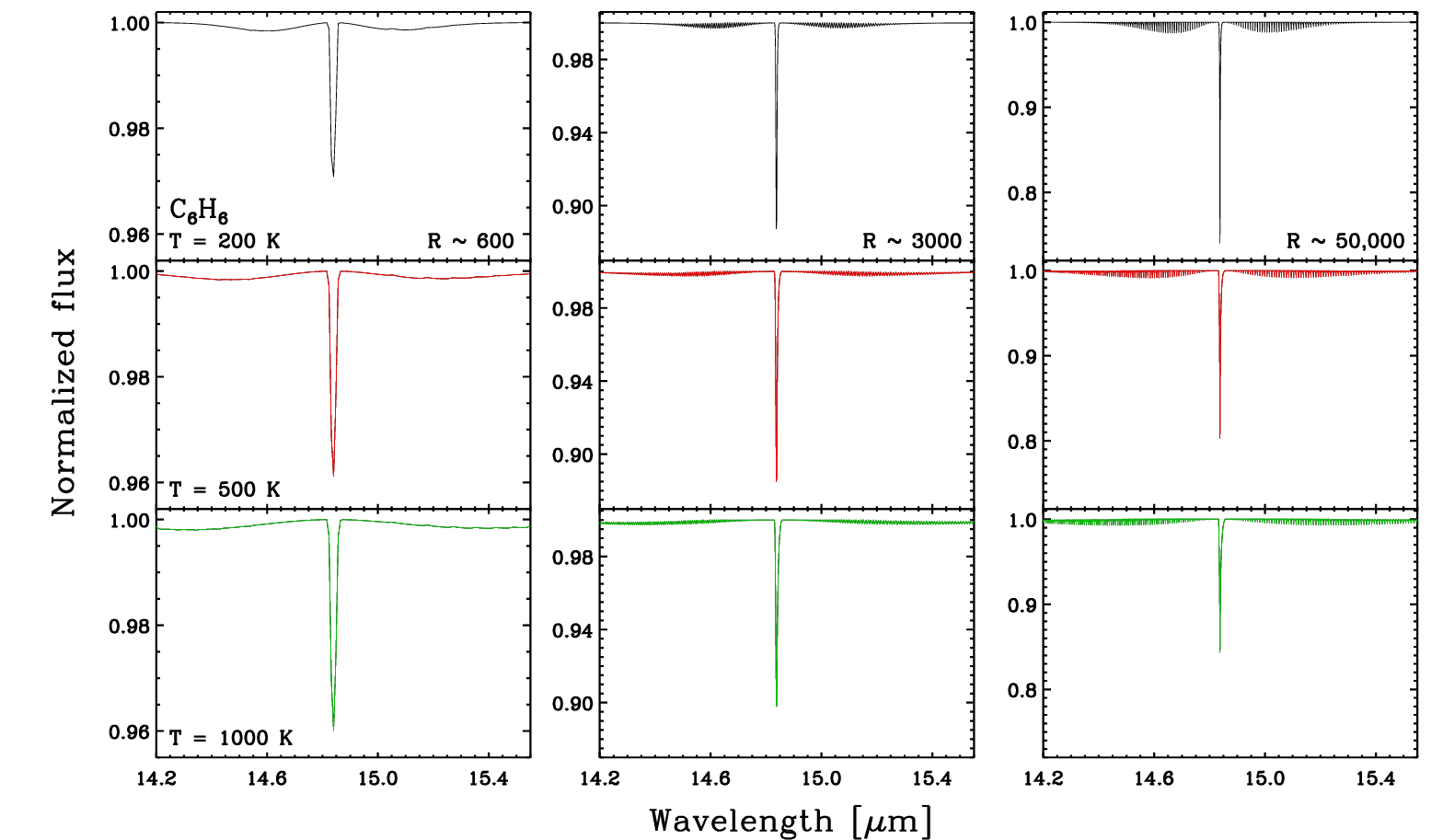}
 \caption{The synthetic spectrum of C$_6$H$_6$ at a column density of 1.0$^.$10$^{16}$ cm$^{-2}$, 
excitation temperatures of 200 (top),
   500 (middle) and 1000 K (bottom), and spectral resolving powers of
   600 (left), 3000 (middle) and 50,000 (right).
}
 \label{fig:c6h6_plot_tot}
}
\end{figure*} 

 \begin{figure*}
\centering
{
 \includegraphics[width=160mm, angle=0.0]{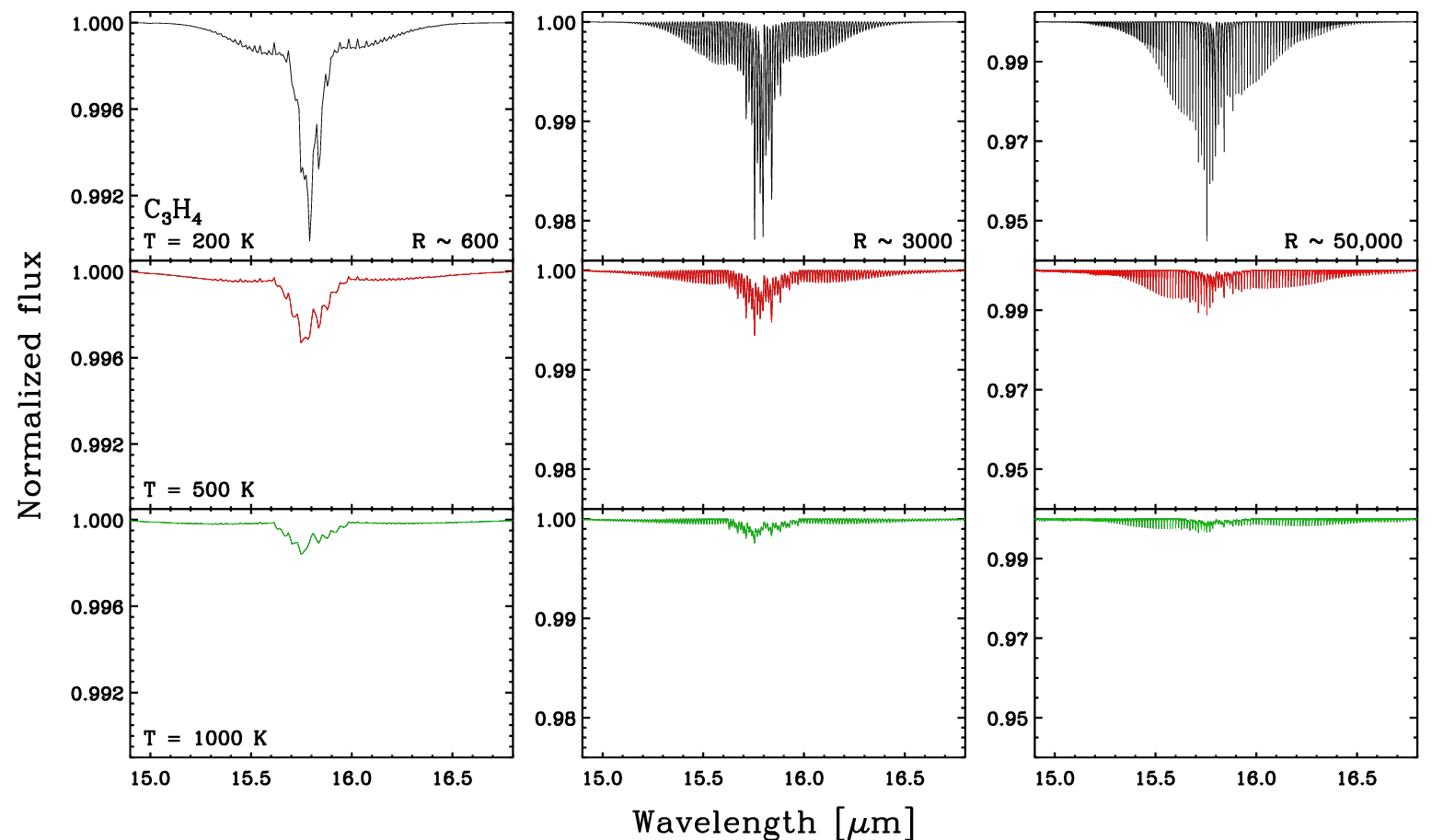}
 \caption{The synthetic spectrum of C$_3$H$_4$ at a column density of 1.0$^.$10$^{16}$ cm$^{-2}$, 
excitation temperatures of 200 (top),
   500 (middle) and 1000 K (bottom), and spectral resolving powers of
   600 (left), 3000 (middle) and 50,000 (right).}
 \label{fig:c3h4_plot_tot}
}
\end{figure*} 

\clearpage

 \begin{figure*}
\centering
{
 \includegraphics[width=160mm, angle=0.0]{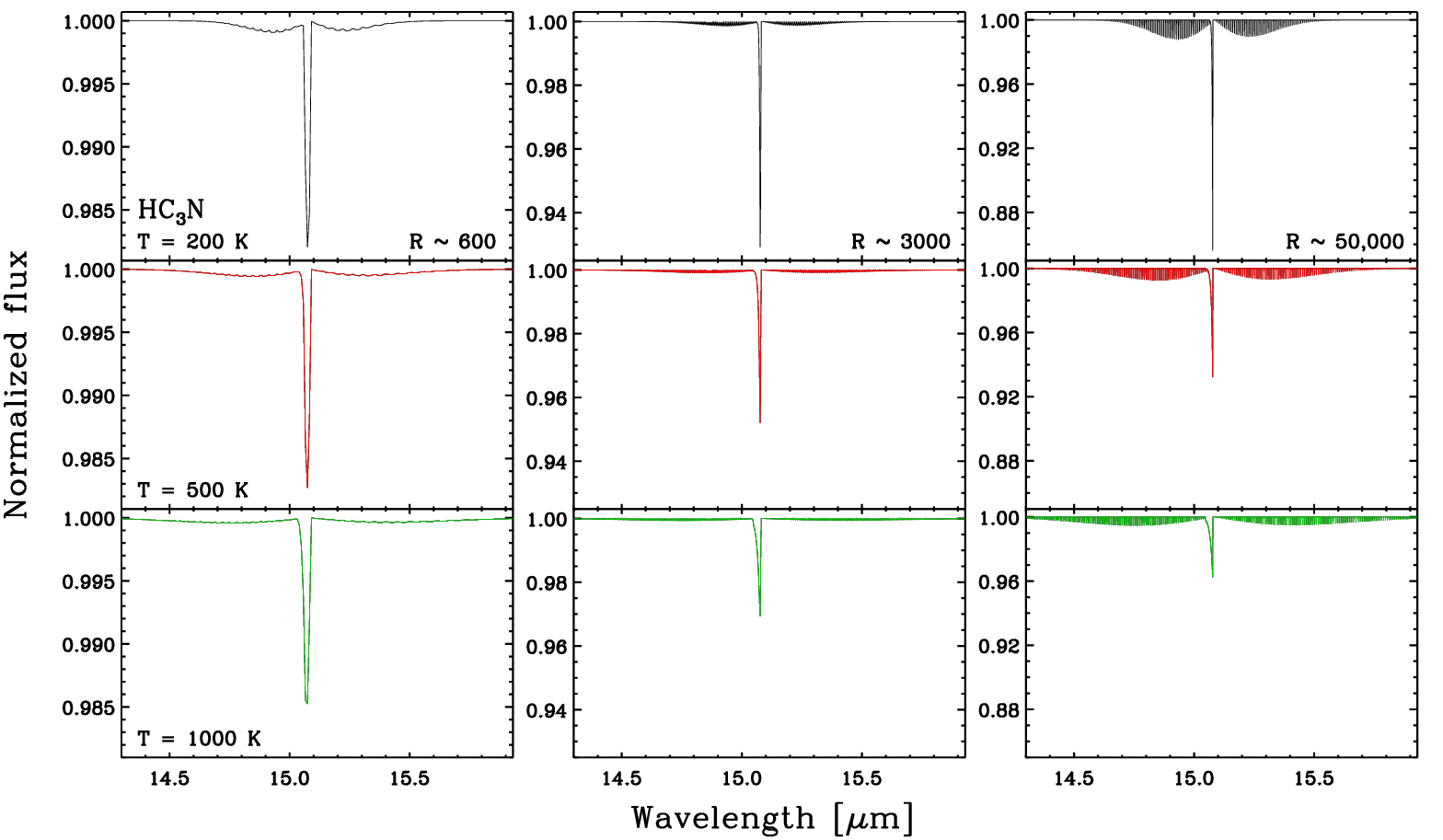}
 \caption{The synthetic spectrum of HC$_3$N at a column density of 1.0$^.$10$^{16}$ cm$^{-2}$, excitation temperatures of 200 (top),
   500 (middle) and 1000 K (bottom), and spectral resolving powers of
   600 (left), 3000 (middle) and 50,000 (right).
}
 \label{fig:hc3n_plot_tot}
}
\end{figure*} 

 \begin{figure*}
\centering
{
 \includegraphics[width=160mm, angle=0.0]{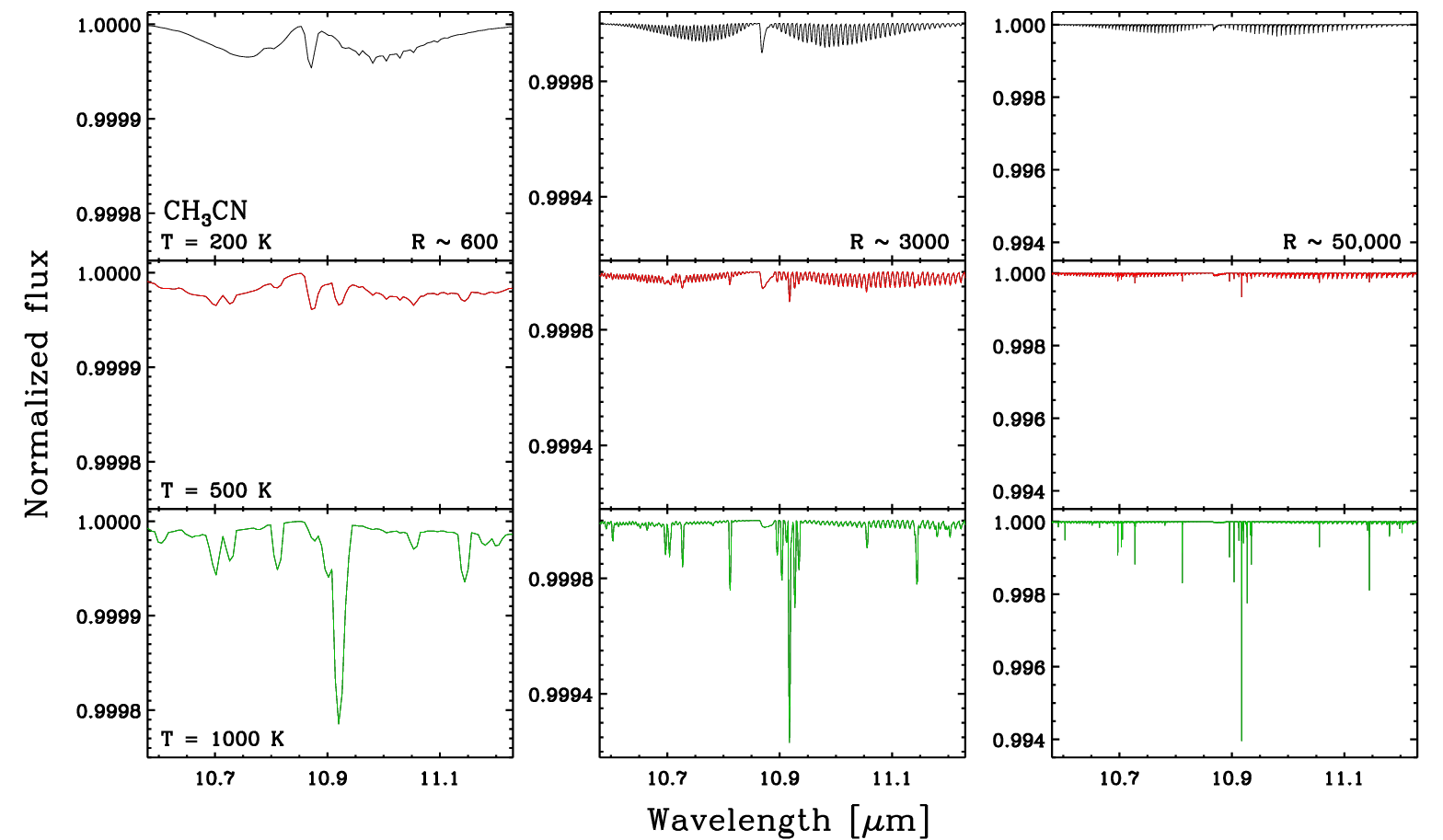}
 \caption{The synthetic spectrum of CH$_3$CN at a column density of 1.0$^.$10$^{16}$ cm$^{-2}$, excitation temperatures of 200 (top),
   500 (middle) and 1000 K (bottom), and spectral resolving powers of
   600 (left), 3000 (middle) and 50,000 (right).
}
 \label{fig:ch3cn_plot_tot}
}
\end{figure*} 

\clearpage

 \end{document}